\documentclass[a4paper,11pt]{article}
\pdfoutput=1
\usepackage{jheppub}
\usepackage[T1]{fontenc}
\usepackage[inline]{enumitem}

\usepackage{alphalph}
\usepackage{etoolbox}

\AtBeginDocument{%
	\AtBeginEnvironment{subequations}{%
	}
}

\title{\boldmath DDF operators in the Hybrid Formalism}

\author{Kiarash Naderi}

\affiliation{Institute for Theoretical Physics, ETH Zurich\\
	8093 Zurich, Switzerland}

\emailAdd{knaderi@phys.ethz.ch}

\abstract{String theory with one unit of NS-NS flux on $\text{AdS}_3 \times \text{S}^3 \times \mathbb{T}^4$ has been recently realised to be the exact dual of the symmetric orbifold of $\mathbb{T}^4$. In the hybrid formalism of Berkovits, Vafa and Witten and using the free-field realisation of $\mathfrak{psu}(1,1|2)_1$, the DDF operators of the free bosons and the free fermions of $\mathbb{T}^4$ are constructed. It is shown that these DDF operators reproduce the whole spectrum of the space-time theory on the world-sheet and therefore provide a derivation of the dual spectrum from first-principles.}

\begin{document} 
	\maketitle
	\flushbottom
	\section{Introduction}
	Recently a concrete realisation of AdS/CFT \cite{Maldacena:1997re} has been found in the tensionless limit. It has been realised that string theory on $\text{AdS}_3 \times \text{S}^3 \times \mathbb{T}^4$ with one unit of NS-NS flux is the exact dual to the symmetric orbifold of $\mathbb{T}^4$ in the large $N$ limit \cite{Gaberdiel:2017oqg,Ferreira:2017pgt,Gaberdiel:2018rqv,Eberhardt:2018ouy,Eberhardt:2019ywk,Dei:2020zui,Eberhardt:2021vsx} (for earlier works on $\text{AdS}_3$, see \cite{Seiberg:1999xz,Argurio:2000tb,Giveon:2005mi,Giribet:2018ada}). In this duality, one has enough analytical control on both sides of the duality, see e.g.\ \cite{Eberhardt:2018ouy,Eberhardt:2019ywk,Dei:2020zui,Eberhardt:2020akk,Knighton:2020kuh,Gaberdiel:2021kkp} for several successful checks. It makes this theory a special laboratory for studying different aspects of $\text{AdS}/\text{CFT}$. In fact, recently there has been a generalisation to $\text{AdS}_5$, see \cite{Gaberdiel:2021qbb,Gaberdiel:2021jrv,Berkovits:2019ulm,Gaberdiel:2022iot} for developments on $\text{AdS}_5 \times \text{S}^5$.
	
	In the RNS formalism, it is not straightforward to define the world-sheet theory at level $k=1$ because the $\mathfrak{su}(2)$ WZW model is at level $k-2=-1$ which does not make sense \cite{Eberhardt:2018ouy,Gaberdiel:2018rqv,Maldacena:2000hw,Seiberg:1999xz}. Instead, the world-sheet is best described in the hybrid formalism of Berkovits, Vafa and Witten (BVW) \cite{Berkovits:1999im,Eberhardt:2018ouy}. In particular, the world-sheet consists of a WZW model on $\mathfrak{psu}(1,1|2)_k$ which describes $\text{AdS}_3 \times \text{S}^3$ and is well-defined for $k=1$. In addition, there are chiral bosons $\rho$ and $\sigma$, and a topologically twisted theory on $\mathbb{T}^4$. The physical states on the world-sheet are described by a double cohomology of an $\mathcal{N}=4$ algebra \cite{Berkovits:1999im}. In the hybrid formalism, the special feature of the level $k=1$ manifests itself by the fact that only the short multiplets of $\mathfrak{psu}(1,1|2)$ are allowed, see \cite{Eberhardt:2018ouy,Dei:2020zui}. In fact, the $\mathfrak{psu}(1,1|2)_1$ WZW model possesses a free-field description in terms of $4$ ``symplectic bosons'' and $4$ free fermions \cite{Eberhardt:2018ouy,Dei:2020zui}. This makes the world-sheet description essentially a combination of free fields and therefore very much under control. 
	
	The free-fields actually realise $\mathfrak{u}(1,1|2)_1$ rather than $\mathfrak{psu}(1,1|2)_1$. In order to get a representation for $\mathfrak{psu}(1,1|2)_1$, one should effectively gauge a particular null current called $Z$ \cite{Dei:2020zui}. The `extra' $Z$-terms in the $\mathfrak{psu}(1,1|2)_1$ anti-commutators in turn only lead to an $\mathcal{N}=4$ algebra on the world-sheet after gauging $Z$ \cite{Gaberdiel:2022als}. However, it is possible to get a ``free-field hybrid formalism'' that has an exact $\mathcal{N}=4$ algebra that at the same time gauges the current $Z$. This has been worked out in \cite{Gaberdiel:2022als} by introducing additional ghosts that effectively quotient out $Z$.
	
	Although the duality mentioned above has been checked extensively, it would be satisfying to have a matching of the spectrum of both sides of the duality from first-principles. Since at level $k=1$ there is a free-field description, there is a good chance of solving the spectrum explicitly in the hybrid formalism. In \cite{Eberhardt:2018ouy}, the partition functions of both sides of the duality were matched using the independence of the ghost system from the level $k$. Moreover, in \cite{Eberhardt:2019qcl,Giveon:1998ns}, the DDF operators \cite{DelGiudice:1971yjh} for a general level $k$ were found, mostly in the RNS formalism. For various applications in the tensionless limit, for example, calculating the correlation functions of certain space-time states from the world-sheet theory, one needs however to find explicit expressions for the physical states in the hybrid formalism. While the RNS and the hybrid formalisms are related to one another by a set of field redefinitions for general level $k$, the dictionary becomes problematic for $k=1$, e.g.\ the $Q_2$ operator of \cite{Gaberdiel:2022als}, which is an exponential in terms of the RNS variables becomes zero in the free-field description \cite{Dei:2020zui,gerick:thesis,Berkovits:1999im}. The aim of this paper is to alternatively construct the DDF operators of the free bosons and the free fermions of the space-time theory on the world-sheet directly in the hybrid formalism at level $k=1$. DDF operators \cite{DelGiudice:1971yjh} are operators that commute with the physical state conditions and therefore map a physical state to a physical state. Similar to the usual string theory on flat background, one would then expect that the whole spectrum is produced by the consecutive applications of these DDF operators on the ground states, while respecting the level-matching conditions.
	
	There are certain difficulties in writing the DDF operators of the free bosons and the free fermions. In the DDF constructions of \cite{Eberhardt:2019qcl,Giveon:1998ns}, the so-called Wakimoto fields are used. An important difficulty is that it is not straightforward to put the Wakimoto fields and the symplectic bosons together in the same language. Another related complication is as follows: Maldacena and Ooguri showed that the world-sheet contains the so-called spectrally flowed representations \cite{Maldacena:2000hw}. In fact, the physical states, in particular the twisted ground states, are in the spectrally flowed representations \cite{Eberhardt:2018ouy,Dei:2020zui}. If one desires to apply the DDF operators to the twisted ground states in order to get the spectrum, one needs to calculate the OPEs between the Wakimoto fields and the world-sheet vertex operators of the spectrally flowed representations in the free-field description.
	
	As a resolution to these challenges, we ``bosonise'' the symplectic bosons. Namely, we introduce two pairs of bosons $(\phi_i,\kappa_i)$ with $i\in\{1,2\}$ so that each pair bosonises a pair of the symplectic bosons, as the symplectic bosons are two $\beta \gamma$ systems with $\lambda=\frac{1}{2}$ \cite{Lust:1989tj}. This solves the above difficulties: \begin{enumerate*}[label=(\roman*)]
		\item we find simple expressions for the Wakimoto fields in the bosonised system,
		\item we write simple explicit formulae for the twisted ground states.
	\end{enumerate*}
	This allows us to achieve the main goal of this paper: we write the DDF operators associated to the free bosons and the free fermions of the space-time theory in the bosonised language. In particular, we show that they commute with the physical state conditions, and also they turn out to satisfy the correct algebra on the world-sheet.
	
	More specifically, upon consecutive applications of these DDF operators to the twisted ground states, we get the whole spectrum of the space-time theory on the world-sheet. All the calculations in this paper focus only on the left-moving (holomorphic) sector. There is another copy of our construction for the right-moving (anti-holomorphic) sector. Once the level-matching conditions are respected \cite{Eberhardt:2018ouy,Maldacena:2000hw}, our construction shows that the world-sheet contains the whole space-time spectrum, and provides explicitly the world-sheet state associated to a given state on the symmetric orbifold.
	
	Apart from the DDF operators for the bosons and the fermions, we identify another set of DDF operators that were previously of interest in the literature \cite{Eberhardt:2019qcl,Giveon:1998ns}. In particular, we identify the DDF operators associated to the $\mathcal{N}=4$ generators of the space-time theory. We confirm that they in fact satisfy the $\mathcal{N}=4$ algebra of the space-time theory on the physical states. Since one expects that the whole spectrum is generated by the $4$ bosons and the $4$ fermions, the DDF operators of the $\mathcal{N}=4$ algebra should produce the same states as suitable combinations of the DDF operators of the free bosons and the free fermions. As a consistency check, we show that this is indeed true in some examples.
	
	The paper is organised as follows. In Section~\ref{sec:vertex}, we review the hybrid formalism and our conventions, the bosonisation of the symplectic bosons and the realisation of the Wakimoto fields. In Section~\ref{sec:ddf}, we construct the DDF operators of the free bosons and the free fermions. In Section~\ref{sec:n=4}, we write the DDF operators associated to the $\mathcal{N}=4$ generators. We also study their algebra and how they are related to the free bosons and the free fermions. Section~\ref{sec:conclusion} contains our conclusions. There are two appendices that include more technical details: in Appendix~\ref{app:bvw}, we discuss the free-field hybrid formalism at $k=1$ in more details. In Appendix~\ref{app:exactness}, we show that $\mathcal{L}_{-1}$ acting on the ground state is BRST exact in the world-sheet.
	
	\section{The bosonisation of the symplectic bosons} \label{sec:vertex}
	In the hybrid formalism of BVW \cite{Berkovits:1999im}, the world-sheet description of string theory with one unit of NS-NS flux on $\text{AdS}_3 \times \text{S}^3 \times \mathbb{T}^4$ consists of $3$ (anti-)commuting theories: \begin{enumerate*}[label=(\roman*)]
		\item a WZW model on $\mathfrak{psu}(1,1|2)_1$,
		\item bosons $\rho$ and $\sigma$,
		\item a topologically twisted CFT on $\mathbb{T}^4$.
	\end{enumerate*}
	Before starting the main topic of this section, which is the bosonisation of the symplectic bosons and the realisation of the Wakimoto fields, we briefly review the main ingredients of each of the mentioned theories and our conventions for them, and we explain how these theories combine together to describe the world-sheet theory. In Appendix~\ref{app:bvw}, we give complementary details that are necessary for performing calculations in Sections~\ref{sec:ddf} and \ref{sec:n=4}. For a detailed review see \cite{Gaberdiel:2022als,gerick:thesis}. Note that in the following we only focus on the left-moving part of the world-sheet theory and there is a similar construction for the right-moving sector as well.
	
	The WZW model on $\mathfrak{psu}(1,1|2)_1$ is described in terms of the currents $J^a$ and $K^a$ with $a\in\{3,\pm\}$ and the supercurrents $S^{\alpha\beta\gamma}$ with $\alpha,\beta,\gamma\in\{\pm\}$ \cite{Eberhardt:2018ouy}. $J^a$ form the affine Lie algebra $\mathfrak{sl}(2,\mathbb{R})_1$, $K^a$ form $\mathfrak{su}(2)_1$ and $S^{\alpha\beta\gamma}$ are the supercurrents that transform in the bi-spinor representation with respect to the bosonic sub-algebra $\mathfrak{sl}(2,\mathbb{R})_1 \oplus \mathfrak{su}(2)_1$. In fact, the WZW model on $\mathfrak{psu}(1,1|2)_1$ has a free-field realisation in terms of $4$ symplectic bosons ($\xi^{\pm}$ and $\eta^{\pm}$) and $4$ fermions ($\chi^{\pm}$ and $\psi^{\pm}$) \cite{Eberhardt:2018ouy,Dei:2020zui}. The free-fields satisfy the following OPEs
	\begin{equation} \label{eq:sym-ferm-opes}
		\xi^{\alpha}(z) \eta^{\beta}(w) \sim \frac{\epsilon^{\alpha\beta}}{(z-w)} \ , \quad \psi^{\alpha}(z) \chi^{\beta}(w) \sim\frac{\epsilon^{\alpha\beta}}{(z-w)} \ ,
	\end{equation}
	where $\alpha,\beta\in \{\pm\}$, $\epsilon^{+-}=-\epsilon^{-+}=1$ and the other combinations vanish. By the symbol ``$\sim$'' we mean the singular part of the OPE. The zero-mode representation of the symplectic bosons is defined as \cite{Dei:2020zui}
	\begin{subequations} \label{eq:symplectic-rep}
		\begin{align}
			\xi^+_0 \left| m_1,m_2 \right> = \left| m_1 ,m_2+\tfrac{1}{2} \right\rangle \ , \qquad & \eta^+_0 \left| m_1,m_2 \right> =2 m_1 \left| m_1 +\tfrac{1}{2},m_2 \right\rangle \ ,\\
			\xi^-_0 \left| m_1,m_2 \right> = -\left| m_1-\tfrac{1}{2} ,m_2 \right\rangle \ , \qquad &  \eta^-_0 \left| m_1,m_2 \right> =-2 m_2 \left| m_1,m_2-\tfrac{1}{2} \right\rangle \ . 
		\end{align}
	\end{subequations}
	The $\mathfrak{psu}(1,1|2)_1$ currents are realised in terms of bilinears of the symplectic bosons and the free fermions \cite{Eberhardt:2018ouy,Dei:2020zui}, and we write them for completeness
	\begin{subequations} \label{eq:u-currents}
	\begin{align}
		J^3(z)& =-\tfrac{1}{2}(\eta^+ \xi^-)(z)-\tfrac{1}{2}(\eta^- \xi^+)(z) \ , \quad & J^{\pm}(z)=(\eta^{\pm} \xi^{\pm})(z) \ , \\
		K^3(z) & =-\tfrac{1}{2}(\chi^+ \psi^-)(z)-\tfrac{1}{2}(\chi^- \psi^+)(z) \ , \quad &  K^{\pm}(z)=\pm(\chi^{\pm} \psi^{\pm})(z) \ , \\
		S^{\alpha\beta+}(z) & =(\xi^{\alpha} \chi^{\beta})(z) \ , \quad &  S^{\alpha\beta-}(z)=(\eta^{\alpha}\psi^{\beta})(z) \ .
	\end{align}
	There is a subtlety that the symplectic bosons and the free fermions realise $\mathfrak{u}(1,1|2)_1$ instead of $\mathfrak{psu}(1,1|2)_1$. In fact, there are $2$ additional bilinear $\mathfrak{u}(1)$ currents which we denote them by $U$ and $V$
	\begin{align}
		U(z)=-\tfrac{1}{2}(\eta^+ \xi^-)(z)+\tfrac{1}{2}(\eta^- \xi^+)(z) \ , \quad V(z)=-\tfrac{1}{2}(\chi^+ \psi^-)(z)+\tfrac{1}{2}(\chi^- \psi^+)(z) \ .
	\end{align}
	It will prove useful to instead consider the following combinations
	\begin{equation}
		Z=U+V \ , \quad Y=U-V \ .
	\end{equation}
	\end{subequations}
	$Z$ commutes with all the $\mathfrak{psu}(1,1|2)_1$ currents and itself, while $Y$ only commutes with all the bosonic $\mathfrak{psu}(1,1|2)_1$ currents and itself. In particular, $Y$ does not commute with $Z$ and the supercurrents $S^{\alpha\beta\gamma}$.  Our conventions for the $\mathfrak{u}(1,1|2)_1$ currents and their (anti-) commutators is almost identical to \cite[Appendix~A]{Gaberdiel:2021njm}, except that for later convenience we have removed a minus sign in $S^{\alpha\beta-}$, see eqs.~(\ref{eq:u-currents}). This also changes the anti-commutator of the supercurrents as follows ($k=1$)
	\begin{align} \label{eq:psu-supercurrents-anti}
		\{S^{\alpha\beta\gamma}_n,S^{\mu\nu\rho}_m\}&=-k n \epsilon^{\alpha\mu} \epsilon^{\beta\nu} \epsilon^{\gamma\rho} \delta_{n+m,0}+\epsilon^{\beta\nu} \epsilon^{\gamma\rho} c_a \sigma_a^{\alpha\mu} J^a_{n+m}\\&-\epsilon^{\alpha\mu}\epsilon^{\gamma\rho} \sigma_a^{\beta\nu} K^a_{n+m}-\epsilon^{\alpha\mu}\epsilon^{\beta\nu} \delta^{\gamma,-\rho} Z_{n+m} \ . \notag
	\end{align}
	This anti-commutator in particular shows why the symplectic bosons and the free fermions only realise $\mathfrak{u}(1,1|2)_1$: it is because there is an additional term proportional to $Z$ on the right hand side compared to the $\mathfrak{psu}(1,1|2)_1$ anti-commutation relations. In fact, for getting a free-field realisation for the $\mathfrak{psu}(1,1|2)_1$ we should focus on the subspace that is annihilated by $Z_n$ for $n\geq 0$ \cite{Dei:2020zui}.
	
	\noindent Before we move on to the next part of the world-sheet theory, recall that for the string theory on $\text{AdS}_3$, one should consider the so called spectrally flowed representations \cite{Maldacena:2000hw,Eberhardt:2018ouy,Dei:2020zui}. The spectral flow is an automorphism of the affine Lie algebra which we denote by $\sigma$. The spectrally flowed representations can be defined as follows: for any integer $w$, we define the representation by $F [v]^{\sigma^w} = [\sigma^w(F) v]^{\sigma^w}$ where $F$ is any generator of $\mathfrak{u}(1,1|2)_1$, and $v$ is any vector in the unflowed representation (e.g.\ see eqs.~(\ref{eq:symplectic-rep})). Our conventions for the action of the spectral flow automorphism $\sigma$ on the symplectic bosons and the fermions in eq.~(\ref{eq:sym-ferm-opes}), and also the generators of $\mathfrak{u}(1,1|2)_1$ is also identical to \cite[Appendix~A]{Gaberdiel:2021njm}, and so we do not repeat them here as it is not directly helpful for our discussion.
	
	The chiral bosons $\rho$ and $\sigma$ satisfy
	\begin{equation}
		\rho(z) \rho(w) \sim -\ln(z-w) \ , \quad \sigma(z) \sigma(w) \sim - \ln(z-w) \ .
	\end{equation}
	The usual diffeomorphism $bc$ system with $c=-26$ is realised as $b=e^{-i\sigma}$ and $c=e^{i\sigma}$, while $\rho$ is a combination of different fields related (but not restricted) to the superdiffeomorphism $\beta\gamma$ system with $c=11$ \cite{Berkovits:1999im,gerick:thesis}, for an explicit formula see e.g.\ \cite[eq.~(C.10) and Appendix~D]{Gaberdiel:2021njm}. The stress-tensor of $\rho$ and $\sigma$ is \cite{Berkovits:1999im,Gaberdiel:2022als}
	\begin{equation}
		T_{\rho\sigma} = -\frac{1}{2} (\partial \rho)^2 - \frac{1}{2} (\partial \sigma)^2 + \frac{3}{2} \partial^2(\rho+i\sigma) \ . 
	\end{equation}

	The topologically twisted $\mathbb{T}^4$ is described in terms of $4$ bosons $\partial X^j$ and $\partial \bar{X}^j$ with $j\in\{1,2\}$ and $4$ free fermions $\psi^j$ and $\bar{\psi}^j$ with $j\in\{1,2\}$. They satisfy the following OPEs
	\begin{equation}
		\partial X^j(z) \partial \bar{X}^k(w) \sim \frac{\delta^{jk}}{(z-w)^2} \ , \quad \psi^j(z) \bar{\psi}^k(w) \sim \frac{\delta^{jk}}{(z-w)} \ .
	\end{equation}
	In the hybrid formalism, it is useful to bosonise the fermions $\psi^j$ and $\bar{\psi}^j$
	\begin{equation}
	\psi^j = e^{iH^j} \ , \quad \bar{\psi}^j=e^{-iH^j} \ ,
	\end{equation}
	with
	\begin{equation}
		H^j(z) H^k(w) \sim -\delta^{jk} \ln(z-w) \ .
	\end{equation}
	We also denote $H=H^1+H^2$. The bosons and fermions form a topologically twisted $\mathcal{N}=4$ superconformal algebra with $c=6$ in terms of their bilinears. We summarise the fields that appear in the topologically twisted $\mathbb{T}^4$ as follows: $T_C$, the topologically twisted stress-tensor (with no central term), $J_C$ and $J_C^{\pm\pm}$, the R-symmetry generators, and $G^{\pm}_C$ and $\tilde{G}^{\pm}_C$, the supercurrents which make two doublets with respect to the R-symmetry currents. We follow exactly the same conventions as \cite[Appendix~B]{Gaberdiel:2021njm} for the generators of the topologically twisted $\mathcal{N}=4$ of $\mathbb{T}^4$ and so we do not rewrite them here. The topological twist stands for the transformation that the stress-tensor $T_{\text{untwisted}}$ of the untwisted theory is shifted to $T_C=T_{\text{untwisted}}+\partial J_C$ where $J_C$ is the Cartan generator of the R-symmetry currents. Throughout the paper, the subscript `C' refers to the topologically twisted $\mathbb{T}^4$ on the world-sheet.
	
	All these $3$ different theories together form a topologically twisted $\mathcal{N}=4$ superconformal algebra with $c=6$ on the world-sheet, whose generators are $T$, the stress-tensor, $J$ and $J^{\pm\pm}$, the R-symmetry generators, and $G^{\pm}$ and $\tilde{G}^{\pm}$, the supercurrents. The explicit expressions of these generators are rather long \cite{Berkovits:1999im,Gaberdiel:2022als} and are reviewed in Appendix~\ref{app:bvw}. The key fact is that a physical state $\phi$ is now described by a double cohomology of this world-sheet topologically twisted $\mathcal{N}=4$ superconformal algebra as follows
	\begin{equation} \label{eq:physical}
		G^+_0 \phi=\tilde{G}^+_0 \phi=(J_0-\tfrac{1}{2})\phi=T_0 \phi = 0 \ , \quad \phi \sim \phi + G^+_0 \tilde{G}^+_0 \psi \ .
	\end{equation}
	Note that we have undone the similarity transformation of \cite{Berkovits:1999im} on the fields in \cite{Gaberdiel:2022als}, see eq.~(\ref{eq:similarity}).
	
	Finding solutions to eq.~(\ref{eq:physical}) is essentially the main goal of this paper. DDF operators \cite{DelGiudice:1971yjh} are operators that commute with the operators that define the physical state conditions, i.e.\ $G^+_0$, $\tilde{G}^+_0$, $J_0$ and $T_0$. Therefore, they map a physical state to a physical state and by their consecutive applications one gets a set of physical states. The goal is to find the DDF operators associated to the free bosons and the free fermions of the dual space-time theory, which is the symmetric orbifold of $\mathbb{T}^4$. We will do this in Section~\ref{sec:ddf}.
	
	There is a subtlety that we have partially ignored up to this point. Although the free-field realisation of $\mathfrak{psu}(1,1|2)_1$ simplifies the BRST analysis and the form of the topologically twisted $\mathcal{N}=4$ superconformal generators, by naively rewriting the world-sheet $\mathcal{N}=4$ generators in terms of the free-fields, one does not get a topologically twisted $\mathcal{N}=4$ algebra \cite{Gaberdiel:2022als}. The reason is that the supercurrents of $\mathfrak{psu}(1,1|2)_1$, written in terms of the free-fields, have `extra' terms proportional to $Z$ in their anti-commutators, see eq.~(\ref{eq:psu-supercurrents-anti}). This in turn changes various OPEs and as a consequence they only match with the OPEs of the topologically twisted $\mathcal{N}=4$ algebra up to terms that are proportional to $Z$ \cite{Gaberdiel:2022als}. In spite of the fact that a simple substitution does not give a topologically twisted $\mathcal{N}=4$ algebra, it is still possible to find an exact topologically twisted $\mathcal{N}=4$ algebra (we have also reviewed this construction in Appendix~\ref{app:bvw}) by adding certain ghosts to the world-sheet that gauge the current $Z$, see \cite{Gaberdiel:2022als}. The argument in \cite{Gaberdiel:2022als} implies that the physical states indeed satisfy
	\begin{equation} \label{eq:z-phys}
		Z_n \phi = 0 \ , \quad (n \geq 0) \ .
	\end{equation}
	Finally, in order to consider the space-time states on the world-sheet, we briefly discuss how the world-sheet vertex operators are set up \cite{Dei:2020zui,Eberhardt:2019ywk}. The space-time M\"obius generators are realised on the world-sheet by the global $\mathfrak{sl}(2,\mathbb{R})_1$ subalgebra of $\mathfrak{u}(1,1|2)_1$. More specifically, $J^{a}_0$ with $a\in\{\pm,3\}$ can be identified with $\mathcal{L}_0=J^3_0$, $\mathcal{L}_{-1}=J^+_0$ and $\mathcal{L}_1=J^-_0$ where $\mathcal{L}_n$ are the left-moving Virasoro generators in the space-time. Having this, to the world-sheet vertex operators we associate an additional label that corresponds to the space-time position, i.e.\ for a state $\phi$ in the world-sheet theory, we consider the vertex operator \cite{Dei:2020zui,Eberhardt:2019ywk}
	\begin{equation} \label{eq:vertex-conjugation}
		V(\phi;x,z)=e^{x J^+_0} e^{z L_{-1}} V(\phi;0,0) e^{-z L_{-1}} e^{- x J^+_0} \ ,
	\end{equation}
	where $L_{-1}$ is the world-sheet translation generator, see eq.~(\ref{eq:n=2-t}). We also note that since $[L_{-1},J^+_0]=0$, this definition is not sensitive to the order of the conjugation.
	
	\subsection{The bosonisation}
	In this subsection, we explain the bosonisation of the symplectic bosons. It allows us to: \begin{enumerate*}[label=(\roman*)] 
		\item write an explicit expression for the $w$-spectrally flowed states,
		\item realise the Wakimoto fields in the bosonised system in Section~\ref{sec:wakimoto}, and
		\item find the DDF operators of the free bosons and the free fermions in Section~\ref{sec:ddf}.
	\end{enumerate*}
	
	We introduce two pairs of bosons $(\phi_i,\kappa_i)$ with $i\in \{1,2\}$ which satisfy
	\begin{equation}
		\phi_i(z) \phi_j(w) \sim -\delta_{ij} \ln(z-w) \ , \quad \kappa_i(z) \kappa_j(w) \sim -\delta_{ij} \ln(z-w) \ .
	\end{equation}
	The symplectic bosons are then realised as follows
	\begin{subequations} \label{eq:symp-exp}
		\begin{equation}
			\xi^-=-e^{-\phi_1-i\kappa_1} \ , \quad \eta^+=e^{\phi_1+i\kappa_1} \partial(i\kappa_1) \ ,
		\end{equation}
		\begin{equation}
			\xi^+=e^{\phi_2+i\kappa_2} \ , \quad \eta^-=-e^{-\phi_2-i\kappa_2} \partial(i\kappa_2) \ .
		\end{equation}
	\end{subequations}
	Note that here we are not explicitly writing the so-called cocycle factors, but we are assuming that they are inserted next to different exponentials of the bosons $\phi_i$ and $\kappa_j$ such that they commute with each other \cite{Goddard:1986ts,Lust:1989tj,Burrington:2015mfa}.\footnote{I would like to especially thank Marc-Antoine Fiset and Vit Sriprachyakul for many useful discussions about this.} It can be checked that the symplectic bosons, as realised in eqs.~(\ref{eq:symp-exp}), indeed satisfy the OPEs in eq.~(\ref{eq:sym-ferm-opes}). The stress-tensor of the symplectic bosons is \cite{Gaberdiel:2022als}
	\begin{equation}
		T_{\text{bosons}}=\frac{1}{2} (\eta^+ \partial \xi^-) - \frac{1}{2} (\xi^- \partial \eta^+)+\frac{1}{2} (\xi^+ \partial \eta^-) - \frac{1}{2} (\eta^- \partial \xi^+) \ ,
	\end{equation}
	where we are using the radial normal-ordering here. The symplectic bosons with respect to $T_{\text{bosons}}$ are primary with weights $\frac{1}{2}$. Taking eqs.~(\ref{eq:symp-exp}) and calculating the stress-tensor leads to
	\begin{equation} \label{eq:added-bosons-t}
		T_{\text{bosons}}=\Big(-\frac{1}{2} \sum_{i=1}^{2} \partial \phi_i \partial \phi_i\Big) +\Big(-\frac{1}{2} \sum_{i=1}^{2} \partial \kappa_i \partial \kappa_i\Big) + \frac{1}{2} \partial^2 (i\kappa_1)-\frac{1}{2} \partial^2(i\kappa_2) \ .
	\end{equation}
	In other words, the $\phi_j$'s have no background charge, while the background charges of the \linebreak$\kappa_i$'s are opposite and equal $\pm 1$. More specifically, for a boson with the following stress-tensor
	\begin{equation}
		T_X = -\frac{1}{2} \partial X \partial X + \frac{\Lambda}{2} \partial^2(i X) , \quad X(z) X(w)\sim -\ln(z-w) \ ,
	\end{equation}
	we say that the background charge of $X$ is $\Lambda$, see \cite{Lust:1989tj}. $T_{\text{bosons}}$ has central charge $c=-2$, in agreement with what one expects from $4$ symplectic bosons since each has $c=-\frac{1}{2}$ \cite{Dei:2020zui}.
	
	The $\mathfrak{sl}(2,\mathbb{R})_1 \times \mathfrak{u}(1) \subset \mathfrak{u}(1,1|2)_1$ generators are written in terms of bilinears of the symplectic bosons, see eqs.~(\ref{eq:u-currents}). Rewriting them in the bosonised language, we get
	\begin{subequations} \label{eq:sl2-free}
		\begin{align}
			J^3&=-\frac{1}{2}(\partial \phi_1+\partial \phi_2) \ , \\
			J^+&=e^{\Sigma} \partial(i\kappa_1) \ , \label{eq:j+-bosonised} \\
			J^-&=e^{-\Sigma} \partial(i\kappa_2) \label{eq:j-} \ , \\
			U&=\frac{1}{2}(-\partial \phi_1+\partial \phi_2) \label{eq:gl2-u} \ ,
		\end{align}
	\end{subequations}
	where $\Sigma$ is defined as
	\begin{equation} \label{eq:sigma}
		\Sigma=\phi_1+i\kappa_1+\phi_2+i\kappa_2 \ .
	\end{equation}
	Note that $U$ is an additional current that commutes with all the $\mathfrak{sl}(2,\mathbb{R})_1$ generators.
	
	As we discussed above, the spectrally-flowed representations are an essential part of the string theory on $\text{AdS}_3$ backgrounds \cite{Maldacena:2000hw}. Focusing on the $\text{AdS}_3$ part, we denote the world-sheet vertex operators associated to the states in the $w$-spectrally flowed representation, namely $[\left|m_1,m_2\right>]^{\sigma^w}$ by $V^w_{m_1,m_2}(x,z)$, see \cite[Appendix~A]{Gaberdiel:2021njm} for our conventions on spectral flow. In \cite{Dei:2020zui}, it is shown that the defining property of the $w$-spectrally flowed states lead to the following OPEs
	\begin{subequations} \label{eq:w-opes}
		\begin{align}
			\xi^{+}(z) V^{w}_{m_1,m_2}(0,0) &= z^{-\frac{w+1}{2}} V_{m_1,m_2+\frac{1}{2}}(0,0) + o(z^{-\frac{w+1}{2}}) \ , \\
			\eta^{+}(z) V^{w}_{m_1,m_2}(0,0) &= 2 m_1 z^{-\frac{w+1}{2}} V_{m_1+\frac{1}{2},m_2}(0,0) + o(z^{-\frac{w+1}{2}}) \ , \\
			\xi^{-}(z) V^{w}_{m_1,m_2}(0,0) &= - z^{\frac{w-1}{2}} V_{m_1-\frac{1}{2},m_2}(0,0) + o(z^{\frac{w-1}{2}}) \ , \\
			\eta^{-}(z) V^{w}_{m_1,m_2}(0,0) &= - 2 m_2 z^{\frac{w-1}{2}} V_{m_1,m_2-\frac{1}{2}}(0,0) + o(z^{\frac{w-1}{2}}) \ ,
		\end{align}
	\end{subequations}
	where we mean $\frac{o(f(z))}{f(z)}\rightarrow 0$ as $z \rightarrow 0$. Having the bosonised version of the symplectic bosons, we can now write the vertex operator associated to the $w$-spectrally flowed state $[\left|m_1,m_2\right>]^{\sigma^w}$ as follows
	\begin{equation} \label{eq:vertex}
		V^w_{m_1,m_2}(0,z)=\exp\Big[{(2m_1+\frac{w-1}{2})\phi_1+2m_1 i \kappa_1+(2m_2+\frac{w+1}{2})\phi_2+2m_2 i \kappa_2}\Big](z) \ .
	\end{equation}
	In fact, by a direct computation, we can see that $V^w_{m_1,m_2}(0,z)$, as in eq.~(\ref{eq:vertex}), satisfies eqs.~(\ref{eq:w-opes}) and therefore realises the $w$-spectrally flowed representation. Note that, as we are only interested in the spectrum in this paper, we have set $x=0$ in eq.~(\ref{eq:vertex-conjugation}) for convenience and as a result, the states have taken the simple form of eq.~(\ref{eq:vertex}).\footnote{It seems difficult to derive a formula for generic $x$ and $w$.}
	
	One can do various consistency checks. For example, it can be seen that $V^w_{m_1,m_2}(0,z)$ from eq.~(\ref{eq:vertex}) satisfies
	\begin{align}
		J^3_0 V^w_{m_1,m_2}(0,z)&=(m_1+m_2+\frac{w}{2}) V^w_{m_1,m_2}(0,z) \ ,\\ U_0 V^w_{m_1,m_2}(0,z)&=(m_1-m_2-\frac{1}{2}) V^w_{m_1,m_2}(0,z) \ ,
	\end{align}
	as expected from the spectral flow of the symplectic bosons in the conventions of \cite[Appendix~A]{Gaberdiel:2021njm}. As another check, these states have the following conformal dimensions with respect to $T_{\text{bosons}}$
	\begin{equation} \label{eq:l_n-sl2r}
		h_w=-\frac{w^2+1}{4}-w (m_1+m_2) \ .
	\end{equation}
	The spectral flow of the stress-tensor associated to the bosonic subalgebra $\mathfrak{sl}(2,\mathbb{R})_1$ in the conventions of \cite[Appendix~A]{Gaberdiel:2021njm} is
	\begin{equation} \label{eq:spectral-flow-l_n-sl2r}
		\sigma^w(L_n)=\tilde{L}_n-w \tilde{J}^3_0 - \frac{w^2}{4} \delta_{n,0} \ .
	\end{equation}
	$\tilde{L}_0$ on $\left|m_1,m_2\right>$ gives $-\frac{1}{4}=4\times (-\frac{1}{16})$, as the symplectic bosons are in the R-sector, see \cite[Section~4.2]{Gaberdiel:2018rqv} and \cite[Appendix~A]{Dei:2020zui}. $\tilde{J}^3_0$ gives the eigenvalue $(m_1+m_2)$ for $\left|m_1,m_2\right>$. As a result, $h_w$ from eq.~(\ref{eq:l_n-sl2r}) matches with the spectral flow formula in eq.~(\ref{eq:spectral-flow-l_n-sl2r}).\footnote{Note that here we are only considering the symplectic bosons. If one adds the fermions to get a free-field realisation for $\mathfrak{u}(1,1|2)_1$, $\tilde{L}_0$ on $\left|m_1,m_2\right>$ gives zero, as each fermion in the R-sector gives $(\frac{1}{16})$, see \cite[Appendix~A]{Dei:2020zui}.}
	
	In the following, it is also sometimes useful to bosonise the fermions of $\mathfrak{u}(1,1|2)_1$
	\begin{equation} \label{eq:fermions-bosonisation}
		\chi^+=e^{iq_1} \ , \quad \psi^-=-e^{-iq_1} \ , \quad \psi^+=e^{iq_2} \ , \quad \chi^-=e^{-iq_2} \ ,
	\end{equation}
	where $q_j$ are bosons satisfying
	\begin{equation}
		q_i(z) q_j(w) \sim -\delta_{ij} \ln(z-w) \ .
	\end{equation}
	The fermions in eq.~(\ref{eq:fermions-bosonisation}) satisfy the OPEs in eq.~(\ref{eq:sym-ferm-opes}). The generators of $\mathfrak{u}(2)_1 \subset \mathfrak{u}(1,1|2)_1$ are written in terms of the free fermions $\chi^{\pm}$ and $\psi^{\pm}$ in eqs.~(\ref{eq:u-currents}). Translating them to the bosonised language, we get
	\begin{subequations} \label{eq:u-su2}
		\begin{align}
			K^3&=\frac{1}{2}\partial(i q_1+iq_2) \ , \\
			K^+&=e^{iq_1} e^{iq_2} \ , \\
			K^-&=e^{-iq_2} e^{-iq_1} \ , \\
			V&=\frac{1}{2}\partial(i q_1-iq_2) \ .
		\end{align}
	\end{subequations}
	Note that $V$ commutes with all the $\mathfrak{su}(2)_1$ generators. Here we again do not explicitly write the cocycle factors, but they are inserted next to $e^{i n q_1}$ and $e^{i m q_2}$ such that they anti-commute for odd $n$ and $m$ \cite{Lust:1989tj,Burrington:2015mfa}. An expression for $Z$ and $Y$ is (see eqs.~(\ref{eq:u-currents}))
	\begin{subequations}
		\begin{equation} \label{eq:z}
			Z=\frac{1}{2}(-\partial \phi_1+\partial \phi_2)+\frac{1}{2}\partial(i q_1-iq_2) \ ,
		\end{equation}
		\begin{equation} \label{eq:y}
			Y=\frac{1}{2}(-\partial \phi_1+\partial \phi_2)-\frac{1}{2}\partial(i q_1-iq_2) \ .
		\end{equation}
	\end{subequations}
	\subsection{The Wakimoto fields} \label{sec:wakimoto}
	Using the bosonisation of the symplectic bosons, it is possible to realise the Wakimoto fields. The Wakimoto representation of $\mathfrak{sl}(2,\mathbb{R})_1$ consists of a pair of commuting ghosts $(\beta,\gamma)$ with weights $(1,0)$, and a boson $\partial \Phi$ of weight $1$ with the following OPEs \cite{Giveon:1998ns,Eberhardt:2019qcl}
	\begin{equation} \label{eq:wakimoto-opes}
		\beta(z) \gamma(w) \sim \frac{-1}{(z-w)} \ , \quad \partial \Phi(z) \partial \Phi(w)\sim \frac{-1}{(z-w)^2} \ ,
	\end{equation}
	where the other OPEs are trivial. The $\mathfrak{sl}(2,\mathbb{R})_1$ generators are written as follows \cite{Eberhardt:2019qcl}
	\begin{subequations}
		\begin{equation}
			J^+=\beta \ ,
		\end{equation}
		\begin{equation}
			J^3=(\beta\gamma)+\frac{i}{\sqrt{2}} \partial \Phi \ ,
		\end{equation}
		\begin{equation} \label{eq:wakimoto-j-}
			J^-=i\sqrt{2}(\partial \Phi \gamma)+(\gamma(\beta \gamma))-\partial \gamma \ .
		\end{equation}
	\end{subequations}
	We have taken this particular normal-ordering in $J^-$ in order to get the correct OPEs. This normal-ordering can be also written as
	\begin{equation} \label{eq:beta-gamma-identity}
		(\gamma(\beta \gamma))-\partial \gamma = ((\beta\gamma)\gamma) \ .
	\end{equation}
	Now we write $(\beta,\gamma)$ and $\partial \Phi$ in terms of the bosons that we have introduced, namely $(\phi_i,\kappa_i)$ with $i\in \{1,2\}$. The first one is easy: $\beta=J^+=e^{\Sigma} \partial(i\kappa_1)$. $\gamma$ is a primary field with weight $0$ such that it has the correct OPE with $\beta$ as in eq.~(\ref{eq:wakimoto-opes}). We set\footnote{Here we are only focusing on the holomorphic parts of $(\beta,\gamma)$ and $\partial \Phi$. It is expected that $\gamma$ and $\partial \Phi$ are holomorphic only near the boundary of $\text{AdS}_3$ \cite{Eberhardt:2019ywk}, see also the discussion at the end of this section.}
	\begin{equation} \label{eq:gamma}
		\gamma=e^{-\Sigma} \ .
	\end{equation}
	It can be checked that it is primary and has a vanishing conformal dimension with respect to $T_{\text{bosons}}$, see eq.~(\ref{eq:added-bosons-t}). Moreover, it has the correct OPE with $\beta$
	\begin{equation}
		e^{\Sigma(z)} \partial(i\kappa_1)(z) e^{-\Sigma(w)} \sim \frac{-1}{(z-w)} \ .
	\end{equation}
	We can see that $\beta$ with itself, and $\gamma$ with itself have trivial OPEs. Note that $\beta$ and $\gamma$ also have trivial OPEs with $U$ in eq.~(\ref{eq:gl2-u}). Finally, we can find $\partial \Phi$ using the expression for $J^3$. First we note that
	\begin{equation}
		(\beta \gamma)=-\partial(\phi_1+\phi_2+i\kappa_2) \ .
	\end{equation}
	Now we can read $i \partial \Phi$ as follows
	\begin{equation}
		i \partial \Phi = \frac{1}{\sqrt{2}} \partial(\phi_1+\phi_2+2 i \kappa_2) \ .
	\end{equation}
	It can be checked that $\partial \Phi$ has the correct OPE with itself as in eq.~(\ref{eq:wakimoto-opes}), and that it also has trivial OPEs with $\beta$, $\gamma$ and $U$. Having these expressions for the Wakimoto fields, one should be able to reproduce $J^-$ as in eq.~(\ref{eq:j-}). Indeed, this can be verified using eq.~(\ref{eq:wakimoto-j-}) and the identity in eq.~(\ref{eq:beta-gamma-identity}). The stress-tensor of the Wakimoto fields is \cite{Eberhardt:2019ywk}\footnote{I thank Beat Nairz for pointing out this to me.}
	\begin{equation}
		T_{\text{Wakimoto}}=-(\beta \partial \gamma)-\frac{1}{2} (\partial \Phi)^2 - \frac{\sqrt{2}}{2} \partial^2 (i\Phi) \ .
	\end{equation}
	As we discussed, the Wakimoto fields, realised in terms of the free-fields $(\phi_i,\kappa_i)$, commute with $U$. In fact, considering
	\begin{equation}
		T_{U+\text{Wakimoto}}=T_{\text{Wakimoto}}-U^2 \ ,
	\end{equation}
	it can be shown that $T_{U+\text{Wakimoto}}=T_{\text{bosons}}$ using the expressions for $\beta$, $\gamma$, $\partial \Phi$ and $U$.
	
	There is a subtlety that we discuss now. Although our free-fields are well-defined and our construction works, one should be careful in using such expressions inside the correlation functions, as we now explain. In fact, by the bosonisation, we are enlarging the symplectic boson theory: for example, the following fields
	\begin{equation}
		\zeta^+ = e^{\phi_1+i\kappa_1} \ , \quad \zeta^-=e^{-\phi_2-i\kappa_2} \ ,
	\end{equation}
	are well-defined in our construction but they are absent in the symplectic boson theory. The fields $\zeta^{\pm}$ appear in the expressions for $\gamma = - (\xi^- \zeta^-)$ and $\gamma^{-1}=(\xi^{+} \zeta^+)$. In \cite{Eberhardt:2019ywk}, it is discussed that (under certain conditions) the correlation function of several spectrally flowed states with the field $\gamma$ inserted equals the associated covering map times the same correlation function. However, as pointed out in \cite{Eberhardt:2019ywk}, covering maps generically have poles away from the positions of the vertex operators and one should cure this issue. In the same paper, it is argued, for example by the charge conservation, that certain `secret representations' should be inserted exactly at the poles of the covering maps in the correlation functions. The same is also true in our bosonised theory. In order to see that our analysis nevertheless works, we note that one can treat the fields $\zeta^{\pm}$ as ``auxiliary fields'', i.e.\ treat them as a formal tool for realising the Wakimoto fields. In fact, using the OPEs of $\zeta^{\pm}$ with the symplectic bosons, eq.~(\ref{eq:vertex}) and $(\xi^{\pm} \zeta^{\mp})=\pm 1$, one can show that any states written in terms of $\zeta^{\pm}$ and the symplectic bosons can be written in terms of the symplectic bosons.\footnote{$\zeta^{\alpha}$ and $\xi^{\beta}$ have trivial OPEs within themselves and $(\xi^{\pm} \zeta^{\mp})=\pm 1$. To prove this statement, we use $(\xi^+ \zeta^-)_{-1}=0$ and let it act on $\left|m_1,m_2\right>$ (eq.~(\ref{eq:vertex}) with $w=0$). It follows that $\zeta^-_{-1} \left|m_1,m_2+\tfrac{1}{2}\right>=-\xi^+_{-1}\left|m_1,m_2-\tfrac{1}{2}\right>$. By induction, this allows us to show that any $\zeta^{\alpha}$- or $\xi^{\beta}$-descendant can be written as a $\xi^{\gamma}$-descendant. Similarly, using that $[\eta^{\alpha}_n,\zeta^{\beta}_m]=\delta^{\alpha\gamma}\delta^{\beta\gamma}(\zeta^{\gamma}\zeta^{\gamma})_{n+m}$ and induction, we can see that the above statement holds.} This shows that the actions of the Wakimoto fields on the spectrally flowed representations is contained in the subspace generated by the symplectic bosons. For these reasons, we believe our construction, at least for the purposes related to the spectrum, works properly.
	
	\section{DDF operators of the bosons and the fermions} \label{sec:ddf}
	In this section, we introduce the DDF operators of the free bosons and the free fermions of $\mathbb{T}^4$. We discuss that they indeed commute with all the physical state conditions in eq.~(\ref{eq:physical}) and therefore map a physical state to a physical state. We show that they form the correct space-time mode algebra on the world-sheet.\footnote{The ``mode algebra'' that we refer to is the algebra that one gets from rewriting the OPEs in \cite[Appendix~B]{Gaberdiel:2021njm} in terms of the modes of the generators of the untwisted $\mathbb{T}^4$. Also see Section~\ref{sec:mode-numbering} for a discussion related to the symmetric orbifold of $\mathbb{T}^4$.} The way that one would reproduce the (single-particle) dual spectrum on the world-sheet now is clear: we act by these bosonic and fermionic DDF operators on the physical states that correspond to the $w$-twisted ground states (also the states with non-zero momenta and winding numbers, see footnote~\ref{footnote} below) of the dual symmetric orbifold of $\mathbb{T}^4$. Once the level-matching conditions are respected, as we briefly review them in Section~\ref{sec:mode-numbering}, our construction shows that the whole spectrum of the space-time theory is reproduced from the world-sheet. In other words, we have found a lower bound on the spectrum of the world-sheet theory. Strictly speaking, one needs to find an argument to provide an upper bound for the world-sheet spectrum, see the discussion in Section~\ref{sec:conclusion}.
	
	\subsection{The $w$-twisted ground states}
	In this subsection, we rewrite the world-sheet states corresponding to the $w$-twisted ground states of the symmetric orbifold of $\mathbb{T}^4$ following \cite{Dei:2020zui} in the bosonised language of Section~\ref{sec:vertex}. The states depend a bit on whether $w$ is odd or even, see \cite{Dei:2020zui}. For $w$ odd, using \cite[Appendix~A.2]{Dei:2020zui} and eq.~(\ref{eq:vertex}) we find\footnote{\label{footnote}Note that here we are not explicitly writing the world-sheet states corresponding to the states with non-zero momenta and winding numbers on the space-time. Focusing on the left-moving part, we consider the state $\left|\{p_j^L\}_{j=1}^4\right>$ on the space-time with $p_j^L$ the left-moving momenta. The associated world-sheet state is a natural generalisation of eqs.~(\ref{eq:w-odd-p=2}) and (\ref{eq:w-even-p=2}). For example, for the untwisted case, the world-sheet state has the form $V^{w=1}_{m_1,m_2} e^{2\rho+i\sigma+iH} \left|\{p_j^L\}_{j=1}^4\right>$ where $m_1$ and $m_2$ are fixed using the physical state conditions and $\left|\{p_j^L\}_{j=1}^4\right>$ is in the topologically twisted $\mathbb{T}^4$ on the world-sheet. The same holds for the twisted cases.}
	\begin{equation} \label{eq:w-odd-p=2}
		\Omega^{w,P=-2}=e^{\frac{w^2-1}{4w}(\phi_1+\phi_2)-\frac{(w-1)^2}{4w}i\kappa_1-\frac{(w+1)^2}{4w}i\kappa_2} e^{2\rho+i\sigma+iH} \ .
	\end{equation}
	For $w$ even, by \cite{Dei:2020zui}, we have
	\begin{equation} \label{eq:w-even-p=2}
		\Omega^{w,P=-2}_{\pm}=e^{\pm \frac{iq_1+iq_2}{2}}e^{\frac{w}{4}(\phi_1+\phi_2)-\frac{(w-2)}{4}i\kappa_1-\frac{(w+2)}{4}i\kappa_2} e^{2\rho+i\sigma+iH} \ .
	\end{equation}
	By a straightforward calculation, it can be checked that they satisfy the physical state conditions in eq.~(\ref{eq:physical}). Note that the space-time ground state is in the untwisted sector which corresponds to $w=1$ in the conventions that we follow \cite{Dei:2020zui}. Therefore, the space-time ground state on the world-sheet is indeed $\Omega^{w=1,P=-2}$ (or in any other picture $P$, see below). As observed in \cite{Dei:2020zui,Gaberdiel:2021njm} and as we confirm later (see Section~\ref{sec:su2}), the global space-time $\mathfrak{su}(2)_1$ should be identified with the global $\mathfrak{su}(2)_1 \subset \mathfrak{ u}(1,1|2)_1$. In fact, $\Omega^{w,P=-2}$ is a singlet for $w$ odd, and $\Omega^{w,P=-2}_{\pm}$ is in a doublet with respect to the global $\mathfrak{su}(2)_1$ of $\mathfrak{u}(1,1|2)_1$, as expected from the space-time perspective, also see \cite{Dei:2020zui}.
	
	The states in eq.~(\ref{eq:w-odd-p=2}) and eq.~(\ref{eq:w-even-p=2}) are in picture $P=-2$, see Appendix~\ref{app:bvw} for our conventions on picture number. In order to relate some of the states that we find using the DDF operators to the states that were previously identified in \cite{Gaberdiel:2021njm}, we write these states in any picture $P=-2n$ for $n \geq 0$. For $w$ odd we find
	\begin{equation} \label{eq:w-odd-p=2n}
		\Omega^{w,P=-2n}=C_{n,w} e^{-(n-1)(iq_1-iq_2)} e^{\frac{w^2-1}{4w}(\phi_1+\phi_2)-\frac{(w-1)^2}{4w}i\kappa_1-\frac{(w+1)^2}{4w}i\kappa_2-(n-1)\Theta} e^{2n\rho+i\sigma+i n H} \ ,
	\end{equation}
	and for $w$ even we get
	\begin{equation} \label{eq:w-even-p=2n}
		\Omega^{w,P=-2n}_{\pm}=B^{\pm}_{n,w} e^{\pm \frac{iq_1+iq_2}{2}-(n-1)(iq_1-iq_2)}e^{\frac{w}{4}(\phi_1+\phi_2)-\frac{(w-2)}{4}i\kappa_1-\frac{(w+2)}{4}i\kappa_2-(n-1)\Theta} e^{2n\rho+i\sigma+i n H} \ ,
	\end{equation}
	where $\Theta$ is defined in eq.~(\ref{eq:theta}), and $C_{n,w}$ and $B^{\pm}_{n,w}$ are numerical factors depending on $n$ and $w$
	\begin{equation} \label{eq:constant-c}
		C_{n,w}=(-1)^{n-1} w^{1-n} \ .
	\end{equation}
	Here we have set $C_{1,w}=1$, and $C_{n,w}$ for $n\geq 0$ is fixed by the fact that they are related by $P_+$, see eq.~(\ref{eq:picture-raising}).
	
	\subsection{Free bosons} \label{sec:bosons}
	For the ``barred bosons'', motivated by the general constructions of \cite{Eberhardt:2019qcl,Giveon:1998ns}, we write
	\begin{equation} \label{eq:ddf-bosons-1}
		\partial \bar{\mathcal{X}}^j_n = \oint dz \partial \bar{X}^j e^{-n\Sigma} \ ,
	\end{equation}
	where $j\in \{1,2\}$, see Appendix~\ref{app:bvw} for our conventions on the hybrid variables. Here $\partial \bar{X}^j$ are the ``barred bosons'' of the topologically twisted $\mathbb{T}^4$ in the world-sheet theory. As we mentioned above, we follow the same conventions as \cite[Appendix~B]{Gaberdiel:2021njm} for the topologically twisted $\mathbb{T}^4$. Also, note that $e^{-n\Sigma}=\gamma^n$ for $n\geq 0$, see eq.~(\ref{eq:gamma}).
	
	The operator in eq.~(\ref{eq:ddf-bosons-1}) commutes with all the physical state conditions in eq.~(\ref{eq:physical}). Actually, we can show the stronger statement that this operator commutes with \textit{all the modes} of the physical state conditions in eq.~(\ref{eq:physical}). First of all, the integrand is primary of weight $1$ and therefore it commutes with all the modes of the world-sheet stress-tensor, see eq.~(\ref{eq:n=2-t}). It can also be seen that it commutes with $Z_n$, see eq.~(\ref{eq:z}), because $Z$ depends on $\phi_1-\phi_2$ while the DDF operator only depends on $\phi_1+\phi_2$. It also commutes with all the modes of $J_n$ and $\tilde{G}^+_n$, see eq.~(\ref{eq:n=2-j}) and eq.~(\ref{eq:n=4-gtillde+}), because the DDF operator does not have any $\rho$, $\sigma$, or $H^k$ contribution. It remains to see whether it commutes with all the modes of $G^+$, see eq.~(\ref{eq:n=2-g+}). In fact, this is the case mainly because in the BRST current $G^+$, the term $G^+_C$ which is (see \cite[eq.~(B.3) and eq.~(B.8)]{Gaberdiel:2021njm})
	\begin{equation} \label{eq:g+-compact}
		G^+_C = \partial \bar{X}^j e^{i H^j}
	\end{equation}
	has a trivial OPE with the barred bosons. The other terms in $G^+$ can also be seen to commute with $\partial \bar{\mathcal{X}}^j_n$. For example, the term $e^{i\sigma} T$ in $G^+$, as the DDF operator does not have any $\sigma$ contribution, commutes with $\partial \bar{\mathcal{X}}^j_n$ because the integrand is primary of weight $1$.
	
	For the ``unbarred bosons'', since $\partial X^j$ has a non-trivial OPE with $G^+_C$, the corresponding DDF operator does not have only a single term. The DDF operator is instead
	\begin{equation} \label{eq:ddf-bosons-2}
		\partial \mathcal{X}^j_n = \oint dz e^{-n \Sigma} \big[ \partial X^j + n \ e^{\rho-\Theta+iH^j} (\psi^+ \psi^-) \big] \ ,
	\end{equation}
	where $j\in \{1,2\}$ and $\Theta$ is defined in eq.~(\ref{eq:theta}), also see eq.~(\ref{eq:fermions-bosonisation}). Note that the second term is still bosonic although $e^{n\rho}$, $e^{im\sigma}$ and $e^{i k_l H^l}$ are fermionic for odd $n$, $m$ and $k_l$. It can be shown that $\partial \mathcal{X}^j_n$ indeed commutes with all the modes of the physical state conditions in eq.~(\ref{eq:physical}). In fact, the integrand is again primary of weight $1$ and commutes with all the modes $Z_n$, $J_n$ and $\tilde{G}^+_n$. It can also be seen that the OPE of $e^{-\rho} Q$ in $G^+$ (see eq.~(\ref{eq:n=2-g+})) with the second term in eq.~(\ref{eq:ddf-bosons-2}), cancels the OPE of $G^+_C$ in $G^+$ with the first term. This shows that $\partial \mathcal{X}^j_n$ is a DDF operator.
	
	We can also read off the algebra between the DDF operators of the free bosons, as one expects to get the space-time algebra between the free bosons. It can be checked that the only non-zero commutator is
	\begin{equation}
		[\partial \mathcal{X}^i_n,\partial \bar{\mathcal{X}}^j_m]=n \delta^{ij} \mathcal{I} \delta_{n+m,0} \ ,
	\end{equation}
	where $\mathcal{I}=-(\partial \Sigma)_0$. Actually, by a direct calculation, instead of $\mathcal{I} \delta_{n+m,0}$, we get
	\begin{equation} \label{eq:identity}
		\mathcal{I}_{n+m}=-\oint dz e^{-(n+m)\Sigma} \partial \Sigma \ .
	\end{equation}
	They are indeed equal because if $k \neq 0$, we have \cite{Eberhardt:2019qcl}
	\begin{equation} \label{eq:identity-argument}
		\mathcal{I}_k=\frac{1}{k}\oint dz \partial (e^{-k\Sigma}) = 0 \ ,
	\end{equation}
	where we have used the fact that the zero mode of an operator with vanishing weight is zero. So actually for the ``identity operator'' we have
	\begin{equation} \label{eq:id}
		\mathcal{I}_k=\mathcal{I} \delta_{k,0} \ , \quad \mathcal{I}=-(\partial \Sigma)_0 \ .
	\end{equation}
	This shows that we have found the DDF operators associated to the free bosons of the space-time theory. In fact, it can be checked that $\mathcal{I}$ commutes with the DDF operators and all the modes of the physical state conditions, mainly because $\Sigma$ with itself and also with $\Theta$ has trivial OPEs, see eq.~(\ref{eq:sigma}) and eq.~(\ref{eq:theta}). Also, the eigenvalue of the $w$-twisted ground states, see eq.~(\ref{eq:w-odd-p=2}) and eq.~(\ref{eq:w-even-p=2}), is $w$ under $\mathcal{I}$.
	
	As a consistency check, we can compare the states that we get using these DDF operators to the states that are found in \cite{Gaberdiel:2021njm}. To this end, for the barred bosons, we calculate $\partial \bar{\mathcal{X}}^j_{-1} \Omega^{w=1,P=0}$ (see eq.~(\ref{eq:w-odd-p=2n}) and eq.~(\ref{eq:constant-c}) with $w=1$ and $n=0$). The result is
	\begin{equation} \label{eq:bosons-1sthalf-state}
		\partial \bar{\mathcal{X}}^j_{-1} \Omega^{w=1,P=0}=-e^{iq_1-iq_2} e^{2\phi_2+i\kappa_2} e^{i\sigma} \partial \bar{X}^j \ ,
	\end{equation}
	which in the notation of \cite{Dei:2020zui,Gaberdiel:2021njm} is
	\begin{equation}
		[\chi^-_{-1}\chi^-_0 \left|0,\tfrac{1}{2}\right>]^{\sigma} e^{i\sigma} \partial \bar{X}^j \ ,
	\end{equation}
	and hence agrees with \cite[eq.~(3.44)]{Gaberdiel:2021njm}. For the unbarred bosons, we calculate \linebreak$\partial \mathcal{X}^j_{-1} \Omega^{w=1,P=-4}$, see eq.~(\ref{eq:w-odd-p=2n}) and eq.~(\ref{eq:constant-c}) with $n=2$ and $w=1$. The second term in $\partial \mathcal{X}^j_{-1}$ gives zero, but from the first term we have
	\begin{equation} \label{eq:bosons-2ndhalf-state}
		\partial \mathcal{X}^j_{-1} \Omega^{w=1,P=-4}=-(\psi^+ \psi^-) e^{2\phi_1+2i\kappa_1-i\kappa_2} e^{4\rho+i\sigma+2 i H} \partial X^j \ , 
	\end{equation}
	which matches with the unbarred bosons in \cite[eq.~(3.47)]{Gaberdiel:2021njm}.
	
	\subsection{Free fermions} \label{sec:fermions}
	For the fermions, we write in picture $P=-1$
	\begin{equation} \label{eq:ddf-fermions-p=1}
		\Psi^{\alpha,j}_r = \oint dz \psi^{\alpha} e^{\phi_1+i\kappa_1} e^{-(r+\frac{1}{2})\Sigma} e^{\rho+iH^j} \ ,
	\end{equation}
	where $\alpha\in\{\pm\}$ and $j\in\{1,2\}$. Note that these operators are fermionic on the world-sheet. Similar to the free bosons, it is possible to show that these operators (anti-)commute with all the modes of the physical state conditions, mainly because the integrand is primary of weight $1$, commutes with $e^{-\rho} Q$ and has trivial OPEs with $G^+_C$, $\tilde{G}^+$ and $J$. Therefore they are indeed DDF operators. We can calculate their algebra as follows: since the identity operator in eq.~(\ref{eq:id}) is in picture $P=0$ and the fermions are in picture $P=-1$, we apply the picture raising operator $P_+$, defined in eq.~(\ref{eq:picture-raising}), to the fermions to get
	\begin{equation} \label{eq:ddf-fermions-p=-1}
		\tilde{\Psi}^{\alpha,j}_r=f_j \Big( \oint dz \chi^{\alpha} e^{\phi_2+i\kappa_2} e^{-(r+\frac{1}{2})\Sigma} e^{-\rho-i H^k} \partial \Sigma + \oint dz \psi^{\alpha} e^{\phi_1+i\kappa_1} e^{-(r+\frac{1}{2})\Sigma} \partial \bar{X}^k \Big) \ ,
	\end{equation}
	where $k \in \{1,2\}-\{j\}$ with $f_1=1$ and $f_2=-1$. Now it can be seen that they satisfy
	\begin{equation} \label{eq:fermions-anti}
		\{\Psi^{\alpha,j}_r,\tilde{\Psi}^{\beta,l}_s\}=\epsilon^{\alpha\beta} \epsilon^{lj} \mathcal{I} \delta_{r+s,0} \ ,
	\end{equation}
	with $\epsilon^{+-}=-\epsilon^{-+}=1$, $\epsilon^{12}=-\epsilon^{21}=1$ and the other combinations vanish. Note that $\mathcal{I}$, defined in eq.~(\ref{eq:id}), commutes with the DDF operators of the free fermions. Using the algebra that we computed, we can relate the spacetime free fermions as follows to the worldsheet DDF operators
	\begin{equation} \label{eq:fermions-identification}
		\psi^1 \leftrightarrow \Psi^{+,1} \ , \quad \psi^2 \leftrightarrow \Psi^{+,2} \ , \quad \bar{\psi}^1 \leftrightarrow -\Psi^{-,2} \ , \quad \bar{\psi}^2 \leftrightarrow \Psi^{-,1} \ .
	\end{equation}
	As a result, we have indeed found the DDF operators associated to the free fermions. Actually, we can confirm that the DDF operators of the bosons and the fermions commute. Taking the fermions in picture $P=-1$ in eq.~(\ref{eq:ddf-fermions-p=1}), we see that they commute with the barred bosons. The same is true for the first term in eq.~(\ref{eq:ddf-bosons-2}) for the unbarred bosons. The second term in eq.~(\ref{eq:ddf-bosons-2}) also has trivial OPEs with the integrand of the fermionic DDF operators because although $e^{\rho(z)}$ and $e^{\rho(w)}$ gives $\frac{1}{(z-w)}$, the normal-ordering of $(\psi^+\psi^-)$ with $\psi^{\alpha}$ gives zero and so they have trivial OPEs.
	
	As in the case of the free bosons, we can do a sanity check by comparing the states that we get from these DDF operators to the states that were identified in \cite{Gaberdiel:2021njm}. Upon calculating $\Psi^{\alpha,j}_{-\frac{1}{2}} \Omega^{w=1,P=0}$ we get
	\begin{equation}
		\Psi^{\alpha,j}_{-\frac{1}{2}} \Omega^{w=1,P=0} = - \chi^{\alpha} e^{\phi_2} e^{\rho+i\sigma+iH^j} \ ,
	\end{equation}
	which are the same states described in \cite[eq.~(3.38)]{Gaberdiel:2021njm} up to an overall numerical factor, and our identification in eq.~(\ref{eq:fermions-identification}) matches with \cite{Gaberdiel:2021njm}.
	
	\subsection{Mode numbering} \label{sec:mode-numbering}
	In order to show that the DDF operators that we have found really reproduce the (single-cycle twisted) spectrum of the symmetric orbifold theory of $\mathbb{T}^4$, we should be able to see that the mode numbers of the DDF operators, once acting on the $w$-twisted ground states, are fractional integers, see e.g.\ \cite[Section~2.2]{Dei:2019iym}. In other words, for a bosonic DDF operator $F_n$, we should have $n \in \mathbb{Z}/w$, and for a fermionic DDF operator $F_r$, we should have $r+\frac{1}{2}\in \mathbb{Z}/w$ upon acting on the $w$-twisted ground states.
	
	An apparent issue on the world-sheet would be that the fractional modes lead to fractional exponents (for example, in eq.~(\ref{eq:ddf-bosons-1}), $n$ could be a fractional integer), and that might violate the single-valuedness of the contour integrals, see \cite{Eberhardt:2019qcl} for a very similar discussion. However, precisely by requiring the single-valuedness of the contour integrals, we can recover the exact forms of the allowed fractional modes. To see this explicitly, we observe that the integrand of the DDF operators that we have introduced has the form $F e^{-n \Sigma}$ for the bosons and $G e^{-(r+\frac{1}{2})\Sigma}$ for the fermions, where $F$ and $G$ are the fields of Sections~\ref{sec:bosons} and \ref{sec:fermions}. Taking the explicit forms of $F(z)$ and $G(z)$ and calculating the OPEs with the $w$-twisted ground states at `point $t$' (see eq.~(\ref{eq:w-odd-p=2}) and eq.~(\ref{eq:w-even-p=2})), it is straightforward to see that their OPEs consist only of integer exponents in $(z-t)$ whether $w$ is odd or even, and therefore it is single-valued. Then let us consider $e^{-k\Sigma(z)}$ where $k=n$ for the bosons and $k=r+\frac{1}{2}$ for the fermions. Similarly, by calculating the OPE of $e^{-k\Sigma(z)}$ with the $w$-twisted ground states at point $t$, we see that they lead to the following contributions (whether $w$ is odd or even)
	\begin{equation}
		(z-t)^{k w} \ .
	\end{equation}
	This forces us to have $k \in \mathbb{Z}/w$ so that the integral is single-valued. By iterative applications of the `fractional modes' of the DDF operators, one adds fractions of $(\phi_j+i\kappa_j)$ with $j\in\{1,2\}$ to the exponent of the $w$-twisted ground states, and also derivatives of various fields. By calculating their OPEs with the integrand of the DDF operators and requiring the single-valuedness, still one is forced to have $k \in \mathbb{Z}/w$. This shows that for the DDF operators of the bosons and fermions we should have $n \in \mathbb{Z}/w$ and $r+\frac{1}{2} \in \mathbb{Z}/w$, respectively. As a result, in our construction, the fractional modes are well-defined and therefore, we reproduce the spectrum of the symmetric orbifold of $\mathbb{T}^4$.
	
	Before closing this section, we briefly review the level-matching conditions that should be respected on the world-sheet. As we mentioned before, our calculations are only for the left-moving (holomorphic) part of the world-sheet theory. There is a similar construction for the right-moving (anti-holomorphic) sector. The orbifold invariance conditions of the symmetric orbifold of $\mathbb{T}^4$ is as follows (see \cite{Eberhardt:2018ouy,Dijkgraaf:1996xw,Dixon:1986qv,Dei:2019iym}): the states on the left and the right sectors either both come from the bosonic or the fermionic fields, or one of them comes from the bosonic and the other from the fermionic fields. Let us denote the left and the right weights by $h$ and $\bar{h}$, respectively. In the former case, only states with $h-\bar{h}\in \mathbb{Z}$ and in the latter case only states with $h-\bar{h}\in \mathbb{Z}+\frac{1}{2}$ are allowed. From the world-sheet perspective, this condition is a consequence of the fact that the world-sheet theory is the diagonal modular invariant, see \cite{DiFrancesco:1997nk,Eberhardt:2018ouy,Maldacena:2000hw}.
	
	\section{The $\mathcal{N}=4$ generators} \label{sec:n=4}
	Similar to the free bosons and the free fermions, one can also try to find the DDF operators associated to the other fields in the space-time theory, for example, the $\mathcal{N}=4$ generators of $\mathbb{T}^4$ \cite{Eberhardt:2019qcl,Giveon:1998ns}. In the space-time theory, these generators are formed by the normal-ordering of the bosons and the fermions, see e.g.\ \cite[Appendix~B]{Gaberdiel:2021njm}.\footnote{As the world-sheet is in some sense the covering space of the symmetric orbifold on $\mathbb{T}^4$, the world-sheet sees only a single copy of $\mathbb{T}^4$, see \cite{Bertle:2020sgd,Gaberdiel:2022oeu}.} In this section, we directly write their DDF operators and the algebras that they form, and we confirm that is identical to the space-time algebra on the physical states. We also compare them to the normal-ordering of the DDF operators associated to the free bosons and the free fermions in the untwisted sector $w=1$, and we find a perfect agreement between these two, as expected at level $k=1$, see \cite{Eberhardt:2019qcl}. In other words, this supports the idea that the whole spectrum is generated by the DDF operators of the bosons and the fermions that we have written in Section~\ref{sec:ddf}. Therefore, the discussions in this section can be seen as further consistency checks.
	
	\subsection{R-symmetry generators} \label{sec:su2}
	The DDF operators associated to the space-time R-symmetry generators are
	\begin{subequations} \label{eq:ddf-su2}
		\begin{align}
			\mathcal{J}_n&=\oint dz K^3 e^{-n\Sigma} \ , \\
			\mathcal{J}^{\pm\pm}_n&=\oint dz K^{\pm} e^{-n\Sigma} \ .
		\end{align}
	\end{subequations}
	Similar to Section~\ref{sec:ddf}, by calculating the OPEs with $G^+$, $\tilde{G}^+$, $J$ and $T$, it can be shown that $\mathcal{J}^a_n$ commute with all the modes of the physical state conditions and $Z$, mainly because $K^b$ have trivial OPEs with $Q$, see eq.~(\ref{eq:q}). Moreover, they satisfy the following algebra
	\begin{subequations}
		\begin{align}
			[\mathcal{J}^{++}_n,\mathcal{J}^{--}_m]&=2 \mathcal{J}_{n+m}+n \mathcal{I} \delta_{n+m,0} \ , \\
			[\mathcal{J}_n,\mathcal{J}_m]&=\frac{n}{2} \mathcal{I} \delta_{n+m,0} \ , \\
			[\mathcal{J}_n,\mathcal{J}^{\pm\pm}_m]&=\pm \mathcal{J}^{\pm\pm}_{n+m} \ ,
		\end{align}
	\end{subequations}
	where the other commutation relations vanish, and $\mathcal{I}$ is defined in eq.~(\ref{eq:id}). This shows that the complete $\mathfrak{su}(2)_1$ of the space-time theory can be identified with the $\mathfrak{su}(2)_1$ of $\mathfrak{u}(1,1|2)_1$, along the same lines of the general constructions of \cite{Eberhardt:2019qcl,Giveon:1998ns}.
	
	In order to find the corresponding states on the world-sheet, we apply $\mathcal{J}^a_{-1}$ to the ground state $\Omega^{w=1,P=-2}$. The result is
	\begin{subequations} \label{eq:su2-state}
		\begin{align}
			\mathcal{J}_{-1} \Omega^{w=1,P=-2} &= K^{3} e^{\phi_1+i\kappa_1+\phi_2} e^{2\rho+i\sigma+iH} \ , \\
			\mathcal{J}^{\pm\pm}_{-1} \Omega^{w=1,P=-2} &= K^{\pm} e^{\phi_1+i\kappa_1+\phi_2} e^{2\rho+i\sigma+iH} \ ,
		\end{align}
	\end{subequations}
	which agree with the states in \cite{Gaberdiel:2021njm}. In addition to this, we can calculate the state that corresponds to the Sugawara stress-tensor associated to the $\mathfrak{su}(2)_1$ of the space-time theory. The final result is
	\begin{align} \label{eq:sugawara}
		&\tfrac{1}{3} \big[\mathcal{J}_{-1} \mathcal{J}_{-1} + \tfrac{1}{2} (\mathcal{J}^{++}_{-1} \mathcal{J}^{--}_{-1} + \mathcal{J}^{--}_{-1} \mathcal{J}^{++}_{-1}) \big] \Omega^{w=1,P=-2}\\&= \big[ T_{\mathfrak{su}(2)_1} e^{2\Sigma-i\kappa_2} + \frac{1}{4} (\partial^2 \Sigma +(\partial \Sigma)^2) e^{2\Sigma-i\kappa_2} \big] e^{2\rho+i\sigma+iH}\ , \notag
	\end{align}
	where $T_{\mathfrak{su}(2)_1}$ equals
	\begin{equation}
		T_{\mathfrak{su}(2)_1}=\tfrac{1}{3} \big[ K^3 K^3 + \tfrac{1}{2} (K^+ K^-+ K^- K^+) \big] \ .
	\end{equation}
	The first line of \cite[eq.~(3.20)]{Gaberdiel:2021njm}, i.e.\ \cite[eq.~(3.21)]{Gaberdiel:2021njm}, gives the first term in the RHS of eq.~(\ref{eq:sugawara}) once $\beta=\frac{1}{4}$, as is mentioned in \cite{Gaberdiel:2021njm}. The second and the third lines in \cite[eq.~(3.20)]{Gaberdiel:2021njm}, once rewritten in the bosonisation language of Section~\ref{sec:vertex}, give
	\begin{equation}
		\beta (\partial^2 \Sigma +(\partial \Sigma)^2) e^{2\Sigma-i\kappa_2} \ .
	\end{equation}
	Hence, the state in eq.~(\ref{eq:sugawara}) agrees with the state that was derived in \cite{Gaberdiel:2021njm}.
	
	Finally, let us compare $\mathcal{J}^a$ to the currents that are formed by the normal-ordering of the DDF operators of the fermions as follows \cite{Eberhardt:2019qcl}
	\begin{subequations} \label{eq:ddf-su2-t4}
		\begin{align} 
			\mathcal{J}^{++,\mathbb{T}^4}_n&=(\psi^1 \psi^2)_n \ , \\
			\mathcal{J}^{--,\mathbb{T}^4}_n&=(\bar{\psi}^2 \bar{\psi}^1)_n \ , \\
			\mathcal{J}^{\mathbb{T}^4}_n&=\tfrac{1}{2}(\psi^j \bar{\psi}^j)_n \ ,
		\end{align}
	\end{subequations}
	where by $\psi^j$ and $\bar{\psi}^j$ we mean the associated world-sheet DDF operators using the identification spelled out in eq.~(\ref{eq:fermions-identification}). The normal-ordering in eqs.~(\ref{eq:ddf-su2-t4}) is the radial normal-ordering expressed in terms of the modes. In fact, the ground state $\Omega^{w=1,P=-2n}$ (for any $n\geq0$) behaves like a vacuum for the fermions, in the sense that only the negative modes give a non-zero result. Using this, taking both fermions in eqs.~(\ref{eq:ddf-su2-t4}) in $P=-1$, it is straightforward to see that the generators in eqs.~(\ref{eq:ddf-su2-t4}) with the mode number $n=-1$ give the same states as in eqs.~(\ref{eq:su2-state}) upon applying to the ground state $\Omega^{w=1,P=0}$. The only point is that for `manifestly' seeing the relations in the vector space rather than in cohomology (i.e.\ up to BRST exact states), we have written
	\begin{equation}
		\mathcal{J}^{\mathbb{T}^4}_n=\tfrac{1}{2}(\psi^j \bar{\psi}^j)_n=\tfrac{1}{2}(\psi^1\bar{\psi}^1)_n-\tfrac{1}{2}(\bar{\psi}^2 \psi^2)_n \ ,
	\end{equation}
	where we have changed the order of the normal-ordering for the second pair of the fermions with a minus sign, that on the space-time does not change $J$. On the world-sheet this is also correct in cohomology: the anti-commutator between two fermions $\Psi^{\alpha,j}$ is proportional to the identity operator $\mathcal{I}_k$ but in picture $P=-2$. Using eq.~(\ref{eq:picture-ddf-state}) and eq.~(\ref{eq:identity-argument}), we see that it only gives a non-BRST exact state for $k=0$. Therefore, two fermions with negative mode numbers anti-commute in cohomology. As a result, we find an agreement between the actions of the DDF operators in eqs.~(\ref{eq:ddf-su2}), and the actions of the generators in eqs.~(\ref{eq:ddf-su2-t4}) on the ground state.
	\subsection{Supercurrents} \label{sec:super}
	The DDF operators of the supercurrents turn out to be
	\begin{subequations} \label{eq:ddf-supercurrents}
		\begin{equation} \label{eq:g+}
			\mathcal{G}^+_r=-\oint dz \chi^{+} e^{\phi_2+i\kappa_2} e^{-(r+\frac{1}{2})\Sigma} \ ,
		\end{equation}
		\begin{equation} \label{eq:gtilde-}
			\tilde{\mathcal{G}}^-_r=\oint dz \chi^{-} e^{\phi_2+i\kappa_2} e^{-(r+\frac{1}{2})\Sigma} \ ,
		\end{equation}
		\begin{align} \label{eq:g-}
			\mathcal{G}^-_r=&-\oint dz \chi^{-} e^{\phi_2+i\kappa_2} e^{-(r+\frac{1}{2})\Sigma} e^{-\rho} G^-_C +\oint dz \chi^{-} e^{\phi_2+i\kappa_2} e^{-(r+\frac{1}{2})\Sigma} e^{-\rho-i\sigma} \\ &+\oint dz e^{-iq_1}[(r+\tfrac{1}{2})(\partial(iq_2)+\partial(\phi_1+i\kappa_1))-\partial(i\kappa_1)] e^{\phi_1+i\kappa_1} e^{-(r+\frac{1}{2})\Sigma} \ , \notag
		\end{align}
		\begin{align} \label{eq:gtilde+}
			\tilde{\mathcal{G}}^+_r=&-\oint dz  \chi^{+} e^{\phi_2+i\kappa_2} e^{-(r+\frac{1}{2})\Sigma} e^{-\rho} G^-_C +\oint dz \chi^{+} e^{\phi_2+i\kappa_2} e^{-(r+\frac{1}{2})\Sigma} e^{-\rho-i\sigma} \\ &-\oint dz e^{iq_2}[(r+\tfrac{1}{2})(-\partial(iq_1)+\partial(\phi_1+i\kappa_1))-\partial(i\kappa_1)] e^{\phi_1+i\kappa_1} e^{-(r+\frac{1}{2})\Sigma} \ . \notag
		\end{align}
	\end{subequations}	
	For deriving these DDF operators, we have used the `global symmetries' of the $\mathcal{N}=4$ algebra in the hybrid formalism \cite{Berkovits:1999im,Gaberdiel:2022als}, so we briefly discuss the relevant parts. These global symmetries form an $\mathcal{N}=4$ algebra and it consists of the zero modes of the $\mathfrak{psu}(1,1|2)_1$ currents, except for half of the supercurrents. The generators of the global symmetry \linebreak(anti-)commute with all the modes of the physical state conditions, see \cite{Berkovits:1999im,Gaberdiel:2022als}. This ``world-sheet'' global $\mathcal{N}=4$ algebra should correspond to the global $\mathcal{N}=4$ algebra of the ``space-time'' theory, where the latter consists of the wedge modes of the $\mathcal{N}=4$ generators in $\mathbb{T}^4$, or more specifically, $L_{\pm 1}$, $L_0$, $J_0$, $J^{\pm\pm}_0$, $G^{\pm}_{\pm \frac{1}{2}}$ and $\tilde{G}^{\pm}_{\pm \frac{1}{2}}$, see \cite[Appendix~B]{Gaberdiel:2021njm}.\footnote{For the R-symmetry generators and the stress-tensor, the corresponding global symmetry generators match with the DDF operators that we have written, see \cite{Berkovits:1999im,Gaberdiel:2022als} and Sections~\ref{sec:su2} and \ref{sec:stress-tensor}.} For the supercurrents of $\mathbb{T}^4$, this means that $G^{\pm}_{\pm \frac{1}{2}}$ and $\tilde{G}^{\pm}_{\pm \frac{1}{2}}$ should correspond to the global supersymmetries in the hybrid formalism. In our conventions (see Section~\ref{sec:vertex} and Appendix~\ref{app:bvw}), the half of the supercurrents that do commute with the physical state conditions are $S^{\alpha\beta+}_0=(\xi^{\alpha} \chi^{\beta})_0$. We have associated them to the space-time supercurrents $G^+$ and $\tilde{G}^-$, see eqs.~(\ref{eq:ddf-supercurrents}). For the other half of the global supersymmetry, within our conventions, we have \cite{Berkovits:1999im,Gaberdiel:2022als}\footnote{Note that have applied the similarity transformation $R$ in eq.~(\ref{eq:similarity}) to the fields in \cite{Gaberdiel:2022als}.}
	\begin{equation}
		\tilde{S}^{\alpha\beta-}_0=\oint dz \Big( S^{\alpha\beta-}+e^{-\rho-i\sigma}S^{\alpha\beta+}-S^{\alpha\beta+}e^{-\rho} G^-_C \Big) \ ,
	\end{equation}
	where $S^{\alpha\beta-}=(\eta^{\alpha} \psi^{\beta})$, see Section~\ref{sec:vertex}. Similarly, we have associated them to the space-time supercurrents $\tilde{G}^+$ and $G^-$, see eqs.~(\ref{eq:ddf-supercurrents}). For writing the DDF operators of the supercurrents, for example $\mathcal{G}^{\pm}$, we have used the following commutators that we expect to hold
	\begin{equation} \label{eq:supercurrent-with-k^3-first-half}
		[\mathcal{J}_n,\mathcal{G}^{\pm}_{-\frac{1}{2}}]=\pm \frac{1}{2} \mathcal{G}^{\pm}_{n-\frac{1}{2}} \ ,
	\end{equation}
	where $\mathcal{J}$ is the DDF operator of the Cartan generator of the R-symmetry, see eqs.~(\ref{eq:ddf-su2}). Then $\mathcal{G}^{\pm}_{r}$ would commute with all the modes of the physical state conditions using the Jacobi identity (possibly except $G^+$, see eq.~(\ref{eq:n=2-g+})), because both $\mathcal{J}_n$ and $\mathcal{G}^{\pm}_{-\frac{1}{2}}$ are DDF operators. The case of $G^+$ is different since $\mathcal{G}^-_{-\frac{1}{2}}$ anti-commutes with all the modes of $G^+$ only up to the terms that are proportional to $Z$
	\begin{equation} \label{eq:global-half-z}
	\{\mathcal{G}^-_{-\frac{1}{2}},G^+_m\}=(Z F)_m \ ,
	\end{equation}
	where $F$ is a combination of the $\mathfrak{psu}(1,1|2)_1$ currents, $\rho$ and $\sigma$, see \cite[eq.~(3.17)]{Gaberdiel:2022als}, that commutes with $Z$. In fact, using the Jacobi identity with $G^+_m$, eq.~(\ref{eq:global-half-z}), and because $\mathcal{J}_n$ commutes with all the modes of $G^+$ and $Z$, we can see $\mathcal{G}^-_r$ anti-commutes with all the modes of $G^+$ up to the $Z$-terms that are zero on the physical states.
	
	One can apply the supercurrent DDF operators to the ground state $\Omega^{w=1,P=-2n}$ in an appropriate picture number, see eq.~(\ref{eq:w-odd-p=2n}), to get the corresponding states on the world-sheet. Before doing that, it is convenient to apply $P_+$, see eq.~(\ref{eq:picture-raising}), to eq.~(\ref{eq:g+}) and eq.~(\ref{eq:gtilde-})
	\begin{align}
		\mathcal{G}^{+,P=1}_r&=\oint dz \chi^{+} e^{\phi_2+i\kappa_2} e^{-(r+1/2)\Sigma} e^{-\rho} \tilde{G}^-_C \ , \label{eq:g+-p=-1} \\
		\tilde{\mathcal{G}}^{-,P=1}_r&=-\oint dz \chi^{-} e^{\phi_2+i\kappa_2} e^{-(r+1/2)\Sigma} e^{-\rho} \tilde{G}^-_C \ . \label{eq:gtilde--p=-1}
	\end{align}
	Now by applying these operators with the mode $r=-\frac{3}{2}$ to $\Omega^{w=1,P=0}$ we get
	\begin{equation} \label{eq:super-1sthalf-state}
		\alpha \ (\partial \chi^{\alpha} \chi^+\chi^-) e^{3\phi_2+2i\kappa_2} e^{-\rho+i\sigma} \tilde{G}^-_C \ ,
	\end{equation}
	where $\alpha\in \{ \pm \}$ (also see eq.~(\ref{eq:constant-c})). This agrees, up to an overall numerical factor with \cite[eq.~(3.26)]{Gaberdiel:2021njm} written in the bosonised form of Section~\ref{sec:vertex}. For the other half of the supercurrents, upon applying them to $\Omega^{w=1,P=-6}$, we get
	\begin{equation} \label{eq:super-2ndhalf-state}
		(\partial \psi^{\alpha} \psi^+ \psi^-) e^{3\phi_1+3i\kappa_1-i\kappa_2} e^{5\rho+i\sigma+2iH}\tilde{G}^+_C \ ,
	\end{equation}
	which also agrees, up to an overall numerical factor, with what is found in \cite{Gaberdiel:2021njm}, once translated to the same language, see \cite[eq.~(3.27-28) and eq.~(3.33-34)]{Gaberdiel:2021njm}. Note that we have chosen the picture $P=-6$ so that only the first terms in $\mathcal{G}^-_{-\frac{3}{2}}$ and $\tilde{\mathcal{G}}^+_{-\frac{3}{2}}$ contribute. Similar to Section~\ref{sec:su2}, it is straightforward to show that the following generators
	\begin{subequations} \label{eq:super-t4}
		\begin{align}
			\mathcal{G}^{+,\mathbb{T}^4}_r&= (\partial \bar{\mathcal{X}}^j \psi^j)_r \ , \\
			\mathcal{G}^{-,\mathbb{T}^4}_r&= (\partial \mathcal{X}^j \bar{\psi}^j)_r \ , \\
			\tilde{\mathcal{G}}^{+,\mathbb{T}^4}_r&= -\epsilon_{ij} (\partial \mathcal{X}^i \psi^j)_r \ ,\\
			\tilde{\mathcal{G}}^{-,\mathbb{T}^4}_r&= -\epsilon_{ij} (\partial \bar{\mathcal{X}}^i \bar{\psi}^j)_r \ ,
		\end{align}
	\end{subequations}
	give the same states with the mode number $r=-\frac{3}{2}$ by acting on the ground state $\Omega^{w=1,P=-2n}$ in appropriate picture numbers. In fact, starting from the states that correspond to the barred and unbarred bosons (see eq.~(\ref{eq:bosons-1sthalf-state}) and eq.~(\ref{eq:bosons-2ndhalf-state})), by applying the DDF operators of the fermions in $P=-1$, one can see that they produce the same states. The only point is that for the supercurrents $\mathcal{G}^+$ and $\tilde{\mathcal{G}}^-$, by the direct calculation we get the states in picture $P=-1$: they indeed match with the states in eq.~(\ref{eq:super-1sthalf-state}) in the picture $P=+1$ by acting by $P_+$ on them, see eq.~(\ref{eq:picture-raising}).
	\subsection{Stress-tensor} \label{sec:stress-tensor}
	Following \cite{Eberhardt:2019qcl,Giveon:1998ns}, we write
	\begin{equation} \label{eq:ddf-stress-tensor}
		\mathcal{L}_n=\oint dz \big[ -\frac{1-n^2}{2}\partial(\phi_1+\phi_2) + \frac{n(n-1)}{2} \partial(i\kappa_1)+\frac{n(n+1)}{2}\partial(i\kappa_2)\big] e^{-n \Sigma} \ .
	\end{equation}
	The normal-ordering prescription that we are using here matches with the one explained in \cite{Gaberdiel:2022als} for the free-fields. We note that $\mathcal{L}_0=J^3_0$, $\mathcal{L}_{-1}=J^+_0$ and $\mathcal{L}_{1}=J^-_0$, as mentioned in Section~\ref{sec:vertex}. By calculating the OPEs with the physical state conditions, one can see that $\mathcal{L}_n$ commutes with all the modes of the physical state conditions and $Z$ (mainly because $\mathcal{L}_n$ commutes with $e^{-\rho} Q$ in $G^+$, see eq.~(\ref{eq:n=2-g+}) and also \cite{Giveon:1998ns,Eberhardt:2019qcl}), and therefore, it is a DDF operator \cite{Eberhardt:2019qcl,Giveon:1998ns}.
	
	Now we can apply $\mathcal{L}_{-2}$ on the ground state $\Omega^{w=1,P=-2}$ to get a state that should correspond to the spacetime stress-tensor, see eq.~(\ref{eq:w-odd-p=2}). The calculation is long, but the final result equals $\Phi e^{2\rho+i\sigma+iH}$ where
	\begin{equation}
		\Phi=\frac{1}{2} e^{2\phi_1+2i\kappa_1+2\phi_2+i\kappa_2} \Big( \partial^2(\phi_1+\phi_2+4 i\kappa_1)+ 2\partial(\phi_1+\phi_2+4i\kappa_1)\partial\Sigma\Big) \ .
	\end{equation}
	Taking the state in \cite[eq.~(3.19)]{Gaberdiel:2021njm} and calculating it in the bosonised version of the symplectic bosons, as in Section~\ref{sec:vertex}, gives $(2\alpha) \Phi e^{2\rho+i\sigma+iH}$. Therefore, this shows that we produce the same result if $\alpha=\frac{1}{2}$, as it is indeed the case \cite{Gaberdiel:2021njm}.
	
	\subsection{Further consistency checks} \label{sec:checks}
	Similar to the discussion in Section~\ref{sec:mode-numbering} for the bosons and the fermions, we should be able to show that the fractional modes of the DDF operators are allowed. In fact, the integrand of all the DDF operators that we have written are of the form $F e^{-k\Sigma}$, where $k=n$ and $k=r+\frac{1}{2}$ for the bosonic and the fermionic generators, respectively. By the explicit forms of $F$'s, we can see that they give integer exponents while $e^{-k\Sigma}$ forces us to have the desired fractional modes, namely $k\in \mathbb{Z}/w$ on the $w$-twisted ground states. This shows that the fractional modes of all the DDF operators are indeed well-defined.
	
	Taking the DDF operators that we have written in Section~\ref{sec:ddf} and Section~\ref{sec:n=4}, one can check various (anti-)commutators. The first obvious commutators are with the ``identity'' operator $\mathcal{I}$. In fact, all the DDF operators commute with $\mathcal{I}$, mainly because $\Sigma$ has trivial OPEs with $\phi_j+i\kappa_j$ for $j\in\{1,2\}$. For the second half of the supercurrents in eqs.~(\ref{eq:ddf-supercurrents}), there is a double pole in the OPE of $\partial \Sigma(z)$ with $\partial(i\kappa_1)(t)$ but it vanishes after taking the integral over $z$, see eq.~(\ref{eq:id}). More interestingly, using similar calculations in \cite{Eberhardt:2019qcl,Giveon:1998ns}, it can be shown that
	\begin{equation} \label{eq:l-algebra}
		[\mathcal{L}_n,\mathcal{L}_m]=(n-m)\mathcal{L}_{n+m}+\frac{1}{2}\mathcal{I} \delta_{n+m,0} n(n^2-1) \ ,
	\end{equation}
	where $\mathcal{I}$ is defined in eq.~(\ref{eq:id}). As we discussed, the fractional modes of $\mathcal{L}_n$ are also well-defined and therefore, eq.~(\ref{eq:l-algebra}) indeed produces the correct space-time Virasoro algebra, see \cite[Section~2.2]{Dei:2019iym}. 
	
	The commutators of the DDF operators associated to the Virasoro primary fields on the space-time with $\mathcal{L}_n$ should show that they are Virasoro primary of an appropriate conformal dimension, namely
	\begin{equation} \label{eq:com-with-t}
		[\mathcal{L}_n,\mathcal{F}_m]=[n(h-1)-m] \mathcal{F}_{n+m} \ ,
	\end{equation}
	where $\mathcal{F}$ is one of the DDF operators that we have written (except $\mathcal{L}_n$, see eq.~(\ref{eq:l-algebra})) and $h$ is the weight. For all the DDF operators, it works in a straightforward way except $2$ cases: \begin{enumerate}
		\item The unbarred bosons, see eq.~(\ref{eq:ddf-bosons-2}). In this case, we note that there is a factor of $m$ in the second term of $\partial \mathcal{X}^j_m$. In calculating $[\mathcal{L}_n,\partial \mathcal{X}^j_m]$, the second term gives a factor of $-(n+m)$, which in together give the second term of $-m\partial \mathcal{X}^j_{n+m}$.
		\item The second half of the supercurrents, see eq.~(\ref{eq:g-}) and eq.~(\ref{eq:gtilde+}). This gives the correct commutator, namely eq.~(\ref{eq:com-with-t}) with $h=\frac{3}{2}$, up to a $Z$-term. In fact, after a tedious calculation, for $\mathcal{G}^-_r$ we get
		\begin{equation}
			[\mathcal{L}_n,\mathcal{G}^-_r]=\Big(\frac{n}{2}-r\Big) \mathcal{G}^-_{r+n} + n(n+1) (\mathcal{ZF})_{r+n} \ ,
		\end{equation}
		where we have defined
		\begin{equation}
			(\mathcal{ZF})_{s}=\oint dz Z e^{-iq_1} e^{\phi_1+i\kappa_1} e^{-(s+\frac{1}{2})\Sigma} \ .
		\end{equation}
		This shows that they are indeed primary of weights $\frac{3}{2}$ on the physical states where the $Z$-terms are zero.
	\end{enumerate}
	
	In addition to these, it is a long and tedious calculation to check all the anti-\linebreak commutators between the supercurrents, see eqs.~(\ref{eq:ddf-supercurrents}). The result is that they all exactly match with the space-time anti-commutation relations, except for
	\begin{equation} \label{eq:super-anti-with-z}
		\{\mathcal{G}^+_r,\mathcal{G}^-_s\}=(r^2-\frac{1}{4}) \delta_{r+s,0} \mathcal{I}+(r-s) \mathcal{J}_{r+s}+\mathcal{L}_{r+s}+(r+s+1)\mathcal{Z}_{r+s} \ ,
	\end{equation}
	and the same for $\{\tilde{\mathcal{G}}^+_r,\tilde{\mathcal{G}}^-_s\}$. Here we have defined
	\begin{equation}
		\mathcal{Z}_n = \oint dt Z(t) e^{-n\Sigma(t)} \ .
	\end{equation}
	As we discussed, on the physical states $Z$-terms are zero and therefore the anti-commutation relation in eq.~(\ref{eq:super-anti-with-z}) matches with the corresponding space-time result on the physical states. Moreover, it is also a lengthy but straightforward calculation to see that the \linebreak(anti-)commutators of the supercurrents with the free fermions and the free bosons agree with the space-time theory. Note that the second term in $\partial \mathcal{X}^j_n$ makes it possible to satisfy
	\begin{equation}
		[\mathcal{G}^+_r,\partial \mathcal{X}^j_n]=-n \Psi^{+,j}_{r+n} \ .
	\end{equation}
	Also, one can check the commutators of all the DDF operators with the DDF operators associated to the space-time R-symmetry generators, see eqs.~(\ref{eq:ddf-su2}). Although the calculations in certain cases are tedious, mainly for the unbarred bosons and the second half of the supercurrents, the final result is that all the other (anti-)commutators agree with the space-time theory without any $Z$-term.
	
	Finally, let us compare $\mathcal{L}_n$ to the stress-tensor in terms of the normal-ordering of the DDF operators of the bosons and the fermions in the untwisted sector \cite{Eberhardt:2019qcl}
	\begin{equation} \label{eq:normal-ordered-t4}
		\mathcal{L}^{\mathbb{T}^4}_n=(\partial \mathcal{X}^j \partial \bar{\mathcal{X}}^j)_n - \frac{1}{2} (\bar{\psi}^j \partial \psi^j)_n- \frac{1}{2} (\psi^j \partial \bar{\psi}^j)_n \ ,
	\end{equation}
	see eq.~(\ref{eq:fermions-identification}). We want to show
	\begin{equation} \label{eq:stress-equality}
		\mathcal{L}_{-2} \Omega^{w=1,P=-2n} = \mathcal{L}^{\mathbb{T}^4}_{-2} \Omega^{w=1,P=-2n} + \text{BRST exact terms}
	\end{equation}
	for an appropriate picture number $P=-2n$ that simplifies the analysis. Before doing this, we state the following
	\begin{equation} \label{eq:exactness-l-1}
		\mathcal{L}_{-1} \Omega^{w=1,P=-2} ~\text{is BRST exact (up to $Z$-terms)} \ .
	\end{equation}
	It means that the translation of the ground state is a BRST exact state, in agreement with the space-time expectation \cite{Dei:2019osr}. This statement has been shown in \cite{Dei:2019osr} in the RNS formulation. We show this directly in the free-field hybrid formalism in Appendix~\ref{app:exactness}. We use this result to argue for eq.~(\ref{eq:stress-equality}).
	
	To begin with, as we discussed in Section~\ref{sec:super}, we have \linebreak$\mathcal{G}^-_{-\frac{3}{2}} \Omega^{w=1,P=-6} = \mathcal{G}^{-,\mathbb{T}^4}_{-\frac{3}{2}} \Omega^{w=1,P=-6}$. By applying $\mathcal{G}^{+,P=1}_{-\frac{1}{2}}$ defined in eq.~(\ref{eq:g+-p=-1}), and $\mathcal{G}^{+,\mathbb{T}^4}_{-\frac{1}{2}}$ in eqs.~(\ref{eq:super-t4}) where the fermions are $\tilde{\Psi}^{\alpha,j}$ (see eq.~(\ref{eq:ddf-fermions-p=-1})), after a long computation we get
	\begin{equation} \label{eq:stress-super-equality}
		\mathcal{G}^{+,P=1}_{-\frac{1}{2}} \mathcal{G}^-_{-\frac{3}{2}} \Omega^{w=1,P=-6} = \mathcal{G}^{+,\mathbb{T}^4}_{-\frac{1}{2}} \mathcal{G}^{-,\mathbb{T}^4}_{-\frac{3}{2}} \Omega^{w=1,P=-6} = -e^{-iq_1+iq_2} e^{3(\phi_1+i\kappa_1)+\phi_2} e^{4\rho+i\sigma+2iH} D\ ,
	\end{equation}
	where
	\begin{equation}
		D=\sum_{j=1}^2 \partial X^j \partial \bar{X}^j + \frac{1}{2} \partial^2 f_j + \frac{1}{2} (\partial f_j)^2 \ , \quad f_j=iq_1+\phi_2+i\kappa_2-\rho-iH^j \ .
	\end{equation}
	By a direct calculation, we also see that $\mathcal{G}^{+,P=1}_{-\frac{1}{2}} \Omega^{w=1,P=-6}=\mathcal{G}^{+,\mathbb{T}^4}_{-\frac{1}{2}} \Omega^{w=1,P=-6}=0$. This implies
	\begin{equation} \label{eq:stress-super-anti}
		\{\mathcal{G}^{+,P=1}_{-\frac{1}{2}},\mathcal{G}^-_{-\frac{3}{2}} \} \Omega^{w=1,P=-6} = \{ \mathcal{G}^{+,\mathbb{T}^4}_{-\frac{1}{2}} ,\mathcal{G}^{-,\mathbb{T}^4}_{-\frac{3}{2}} \} \Omega^{w=1,P=-6} \ .
	\end{equation}
	Consider the following difference
	\begin{equation}
		A=\Big[\{\mathcal{G}^{+}_{-\frac{1}{2}},\mathcal{G}^-_{-\frac{3}{2}} \} - \{ \mathcal{G}^{+,\mathbb{T}^4}_{-\frac{1}{2}} ,\mathcal{G}^{-,\mathbb{T}^4}_{-\frac{3}{2}} \} \Big] \Omega^{w=1,P=-6} \ ,
	\end{equation}
	where now $\mathcal{G}^{+}_{-\frac{1}{2}}$ is defined in eqs.~(\ref{eq:ddf-supercurrents}) and in $\mathcal{G}^{+,\mathbb{T}^4}_{-\frac{1}{2}}$ the fermions are $\Psi^{\alpha,j}$ in picture $P=-1$, see eq.~(\ref{eq:ddf-fermions-p=1}). Let us apply $P_+$ on $A$. Using eq.~(\ref{eq:picture-ddf-state}) and eq.~(\ref{eq:stress-super-anti}), we get that up to BRST exact states we have
	\begin{equation}
		P_+ A = 0 \ .
	\end{equation}
	It means that $A$ is zero in a specific picture. Assuming the kernel of $P_+$ is zero in cohomology, this shows that $A$ is BRST exact \cite{Lust:1989tj}.\footnote{However, we have not been able to directly show that the kernel of $P_+$ is zero in cohomology.} This consequently implies that, up to BRST exact terms, we have
	\begin{equation} \label{eq:equity-l}
		\Big(\mathcal{J}_{-2}+\mathcal{L}_{-2}\Big) \Omega^{w=1,P=-6} = \Big(\mathcal{J}^{\mathbb{T}^4}_{-2}+\mathcal{L}^{\mathbb{T}^4}_{-2}\Big) \Omega^{w=1,P=-6} \ ,
	\end{equation}
	see eq.~(\ref{eq:super-anti-with-z}). Here we are ignoring the terms that are proportional to $Z$ since they are zero on the physical states. Using the results of Section~\ref{sec:su2}, up to BRST exact terms, we have
	\begin{equation} \label{eq:su2-equality}
		\mathcal{J}_{-1} \Omega^{w=1,P=-6}=\mathcal{J}^{\mathbb{T}^4}_{-1} \Omega^{w=1,P=-6} \ ,
	\end{equation}
	as we have shown the equality in another picture. More accurately, by applying $P_+$ twice and using eq.~(\ref{eq:picture-ddf-state}), we can see that the difference is zero because of the equality in $P=-2$ (see Section~\ref{sec:su2}). Now by applying $\mathcal{L}_{-1}$ on both sides, using the commutators in eq.~(\ref{eq:com-with-t}) and the statement in (\ref{eq:exactness-l-1}), we can see that
	\begin{equation}
		\mathcal{J}_{-2} \Omega^{w=1,P=-6}=\mathcal{J}^{\mathbb{T}^4}_{-2} \Omega^{w=1,P=-6} + \text{BRST exact terms} \ .
	\end{equation}
	Hence, using eq.~(\ref{eq:equity-l}), we conclude that
	\begin{equation}
		\mathcal{L}_{-2} \Omega^{w=1,P=-6}=\mathcal{L}^{\mathbb{T}^4}_{-2} \Omega^{w=1,P=-6} + \text{BRST exact terms} \ .
	\end{equation}
	
	\section{Conclusions and Remarks} \label{sec:conclusion}
	The aim of this paper was to systematically reproduce the spectrum of the symmetric orbifold of $\mathbb{T}^4$ on the world-sheet at level $k=1$ in the hybrid formalism. In order to do this, we bosonised the symplectic bosons and introduced the DDF operators of the free bosons and the free fermions. By the consecutive applications of these DDF operators to the $w$-twisted ground states while respecting the level-matching conditions, our construction provides a direct derivation of the whole spectrum of the space-time theory on the world-sheet.
	
	Motivated by the previous analysis in \cite{Eberhardt:2019qcl,Giveon:1998ns}, we also wrote the DDF operators associated to the $\mathcal{N}=4$ superconformal generators of the space-time theory in the bosonised language. We discussed that they form exactly the space-time symmetry algebra on the physical states. In fact, the $\mathcal{N}=4$ algebra on the space-time is formed by the normal-ordering of the bosons and the fermions and one expects that this holds on the world-sheet as well. We observed that it is indeed the case in some examples. This provides further evidence supporting that the spectrum is generated by the $4$ bosons and the $4$ fermions at $k=1$, see \cite{Eberhardt:2019qcl}.
	
	Using the explicit form of the DDF operators associated to the bosons and the fermions, and the discussion at the end of Section~\ref{sec:wakimoto}, we can see that they never give rise to any $\eta^{\pm}$ by their consecutive applications to the $w$-twisted ground states. This suggests that perhaps a generic physical state that contains $\eta^{\pm}$ can be written as a state with no $\eta^{\pm}$ plus a BRST exact state. Assuming this, the correlation functions of any number of the physical states on the world-sheet might be only delta-function localised, see \cite{Dei:2020zui}. Of course, this requires further investigation.
	
	We discuss a few possible directions for future research. Our construction provides a `lower bound' for the world-sheet spectrum. One might be able to put an `upper bound' on the spectrum by proving that any physical state is a descendant of the $w$-twisted ground states (and the states with non-zero momenta and winding numbers) by applications of the DDF operators, plus a BRST exact state. In other words, only $4+4$ bosonic and fermionic degrees of freedom are left after imposing the physical state conditions. The analysis of \cite{Eberhardt:2019ywk} shows that the partition functions of the both sides of the duality agree, therefore it provides strong evidence that the upper bound is the same as the lower bound. This essentially shows the no-ghost theorem because the algebra of the DDF operators of the free bosons and the free fermions (which is identical to the algebra in the space-time theory on $\mathbb{T}^4$) gives manifestly non-negative norms. Alternatively, one might hope to find an argument similar to the Goddard and Thorn no-ghost theorem in the hybrid formalism \cite{Bienkowska:1991zs,Goddard:1972iy} to enumerate the space of physical states (see \cite{Hwang:1990aq,Hwang:1991ana,Evans:1998qu} for a no-ghost theorem for $\text{AdS}_3$ in the RNS formalism). In fact, we suspect that working with an $\mathcal{N}=2$ critical description may simplify the analysis \cite{Berkovits:1993xq}. Actually, as we mentioned in the text, the DDF operators that we have found commute with all the modes of the critical $\mathcal{N}=2$ generators up to $Z$-terms, and this suggests that they produce the spectrum even in the critical description. A related question is gaining a better understanding of the relation between the null states of the space-time and the BRST exact states on the world-sheet.
	
	A natural question is whether one can produce the space-time correlators of arbitrary states from the world-sheet in a somewhat manifest way, see \cite{Gaberdiel:2022oeu} for a related work. Another connected direction is to study the $W_{\infty}$ symmetry of the space-time theory \cite{Gaberdiel:2011wb,Gaberdiel:2012uj,Gaberdiel:2015mra}, and see whether one can realise it (manifestly) on the world-sheet. One might be able to relate the world-sheet OPEs to the space-time OPEs to reproduce the underlying symmetries more clearly from the world-sheet. As another question, considering the bosonisation of the symplectic bosons, one can study whether they lead to a more direct approach for calculating the correlation functions rather than using the Ward identities in \cite{Dei:2020zui}.
	
	Furthermore, it would be interesting to study the duality that we have considered away from the ``free-field point'', namely by perturbing the space-time theory and studying the effect on the world-sheet or vice versa. For instance, our analysis may help in determining the world-sheet vertex operators associated to the exactly marginal operators in the space-time. It allows one to study the moduli space and compare the result to the space-time theory \cite{Fiset}, also see \cite{Gaberdiel:2015uca} for the calculation in the space-time.
	
	\acknowledgments
	I would like to thank Matthias Gaberdiel for crucial discussions and his important guidance through this work. I am also grateful for his detailed comments on draft versions of the paper. It is my pleasure to thank Marc-Antoine Fiset and Vit Sriprachyakul for related discussions at an early stage of this work and their comments on the draft. I would like to also thank Bob Knighton and Beat Nairz for useful discussions and comments on the draft. This work was supported by a grant from the Swiss National Science Foundation (SNSF) as well as the NCCR SwissMAP that is also funded by the SNSF.
	
	\appendix
	\section{The hybrid formalism} \label{app:bvw}
	In this appendix, we review some of the details of the hybrid formalism that will be needed for the calculations in Sections~\ref{sec:ddf} and \ref{sec:n=4}. In particular, we review the hybrid formalism of BVW based on \cite{Berkovits:1999im,Gaberdiel:2022als,gerick:thesis} with one unit of NS-NS flux, and also the concept of picture changing. Here we only focus on the left-moving (holomorphic) sector.
	
	As we mentioned in Section~\ref{sec:vertex}, the world-sheet description consists of a WZW model on $\mathfrak{psu}(1,1|2)_1$, chiral bosons $\rho$ and $\sigma$, and a topologically twisted $\mathbb{T}^4$. We reviewed our conventions for each theory in Section~\ref{sec:vertex}. In particular, the $\mathfrak{psu}(1,1|2)_1$ is described in terms of bilinears of the $4$ symplectic bosons ($\xi^{\pm}$ and $\eta^{\pm}$) and the $4$ fermions ($\chi^{\pm}$ and $\psi^{\pm}$). See Section~\ref{sec:vertex} and Appendices~A and B of \cite{Gaberdiel:2021njm} for our conventions on the generators of the $\mathfrak{psu}(1,1|2)_1$ and $\mathbb{T}^4$.
	
	The free-field realisation of $\mathfrak{psu}(1,1|2)_1$, as we reviewed in Section~\ref{sec:vertex}, leads to a simplification for the generators of the world-sheet $\mathcal{N}=2$ superconformal algebra but it does not quite form an $\mathcal{N}=2$ algebra or its topologically twisted version. However, it is possible to add certain ghosts to achieve this, as is done in \cite{Gaberdiel:2022als}. In fact, a pair of anti-commuting ghosts $(b,c)$ and $(b^{\prime},c^{\prime})$, and a commuting ghost $(\beta^{\prime},\gamma^{\prime})$ with weights $(1,0)$ are added to the world-sheet theory that effectively quotient out the current $Z$, see \cite{Gaberdiel:2022als}. The additional ghosts satisfy the following OPEs
	\begin{equation}
		b(z) c(w) \sim \frac{1}{(z-w)} \ , \quad b^{\prime}(z) c^{\prime}(w)\sim\frac{1}{(z-w)} \ , \quad \beta^{\prime}(z) \gamma^{\prime}(w)\sim\frac{1}{(z-w)} \ .
	\end{equation}
	The topologically twisted $\mathcal{N}=2$ generators with $c=6$ take the following form \cite{Berkovits:1999im,Gaberdiel:2022als}
	\begin{subequations} \label{eq:n=2}
		\begin{equation} \label{eq:n=2-t}
			T=T_{\text{int}}-\frac{1}{2}\partial \rho \partial \rho-\frac{1}{2} \partial \sigma \partial \sigma+\frac{3}{2} \partial^2(\rho+i\sigma)+T_C
		\end{equation}
		\begin{equation} \label{eq:n=2-g+}
			G^+=e^{-\rho} Q + e^{i\sigma} T -\partial(e^{i\sigma} \partial(\rho+iH))+G^+_C \ ,
		\end{equation}
		\begin{equation} \label{eq:n=2-g-}
			G^-=e^{-i\sigma} \ ,
		\end{equation}
		\begin{equation} \label{eq:n=2-j}
			J=\tfrac{1}{2}\partial(\rho+i\sigma+iH) \ ,
		\end{equation}
	\end{subequations}
	with the following definitions \cite{Dei:2020zui,Gaberdiel:2022als}
	\begin{subequations}
		\begin{equation} \label{eq:q}
			Q=(\chi^+\chi^-)(\xi^+\partial \xi^- - \xi^- \partial \xi^+)= e^{iq_1-iq_2} e^{\Theta} \partial \Sigma \ ,
		\end{equation}
		\begin{equation} \label{eq:theta}
			\Theta = -\phi_1-i\kappa_1+\phi_2+i\kappa_2 \ .
		\end{equation}
		\begin{equation} \label{eq:tint}
			T_{\text{int}}=T_{\mathfrak{u}(1,1|2)_1}+T_{bc}+T_{b^{\prime}c^{\prime}}+T_{\beta^{\prime}\gamma^{\prime}}+\partial(\gamma^{\prime} Z) \ ,
		\end{equation}
		\begin{equation}
			T_{bc}=((\partial c) b) \ , \quad T_{b^{\prime}c^{\prime}}=((\partial c^{\prime}) b^{\prime}) \ , \quad T_{\beta^{\prime}\gamma^{\prime}}=((\partial \gamma^{\prime})\beta^{\prime}) \ .
		\end{equation}
	\end{subequations}
	A few explanations are in order: note that the $Q_2$ term in \cite[eq.~(2.13b)]{Gaberdiel:2022als}, as is mentioned in \cite{Dei:2020zui,Gaberdiel:2022als}, is zero in the free-field realisation. $T_{\mathfrak{u}(1,1|2)_1}$ is the stress-tensor of $\mathfrak{u}(1,1|2)_1$, see \cite[eq.~(3.2)]{Gaberdiel:2022als}, while the symplectic bosons part is now replaced by $T_{\text{bosons}}$, see eq.~(\ref{eq:added-bosons-t}). The subscript `C' indicates the $\mathcal{N}=4$ generators of the topologically twisted $\mathbb{T}^4$, see \cite[Appendix~B]{Gaberdiel:2021njm} and eq.~(\ref{eq:g+-compact}). As we mentioned in the main text, the boson $H=H^1+H^2$ comes from the bosonisation of the fermions of the compact theory $\mathbb{T}^4$, namely
	\begin{equation}
		\psi^j = e^{iH^j} \ , \quad \bar{\psi}^j=e^{-iH^j} \ , \quad J_C = \tfrac{1}{2} \partial(iH) \ ,
	\end{equation}
	with $j\in\{1,2\}$ and
	\begin{equation}
		H^j(z) H^k(w) \sim -\delta^{jk} \ln(z-w) \ .
	\end{equation}
	We note that the generators in eqs.~(\ref{eq:n=2}) are the generators of the $\mathcal{N}=2$ topologically twisted algebra of \cite{Berkovits:1999im} \textit{before} the similarity transformation $R$ defined as \cite{Berkovits:1999im}
	\begin{equation} \label{eq:similarity}
		R = \oint e^{i\sigma} G^-_C \ . 
	\end{equation}
	In other words, the generators in eqs.~(\ref{eq:n=2}) are the generators in \cite[eqs.~(4.2)]{Gaberdiel:2022als} with the similarity transformation $R$ acted on them. It was already predicted in \cite{Gaberdiel:2021njm} that the generators before the similarity transformation of \cite{Berkovits:1999im} might simplify the BRST analysis, as we also observed in this paper. We note that the normal-ordering prescription here matches with \cite{Gaberdiel:2022als}.
	
	The $\mathcal{N}=4$ generators are then constructed out of the $\mathcal{N}=2$ fields. In fact, the $\mathfrak{u}(1)$ generated by $J$ can be enhanced to an affine $\mathfrak{su}(2)_1$. More specifically, the $\mathfrak{su}(2)_1$ generators are $J$ and $J^{\pm\pm}$ with $J$ defined in eq.~(\ref{eq:n=2-j}) and
	\begin{equation} \label{eq:n=4-js}
		J^{\pm\pm}=e^{\pm(\rho+i\sigma+iH)} \ .
	\end{equation}
	The additional $\mathcal{N}=4$ supercurrents then take the following form
	\begin{subequations}
		\begin{equation} \label{eq:n=4-gtillde+}
			\tilde{G}^+=e^{\rho+i H} \ ,
		\end{equation}
		\begin{equation}
			\tilde{G}^-= e^{-2\rho-i\sigma-iH} Q -e^{-\rho-iH} T -e^{-\rho-i\sigma} \tilde{G}^-_C + e^{-\rho-iH}[i\partial \sigma \partial(\rho+iH)+\partial^2(\rho+iH)] \ .
		\end{equation}
	\end{subequations}
	
	Now we discuss the picture number and the picture raising operator. The picture number is defined as the eigenvalue of the operator
	\begin{equation}
		P=-\partial(\phi+i\kappa) \ ,
	\end{equation}
	where $\phi$ and $\kappa$ are the bosons that appear in the bosonisation of the superdiffeomorphism $\beta\gamma$ system, see e.g.\ \cite[Appendix~D]{Gaberdiel:2021njm}. The $\phi$ and $\kappa$ here should not be confused with the bosons $(\phi_i,\kappa_i)$ with $i\in \{1,2\}$ in the bosonisation of the symplectic bosons in Section~\ref{sec:vertex}. Note that relative to \cite{Gaberdiel:2021njm}, we have changed the definition of $P$.\footnote{In \cite{Gaberdiel:2021njm}, the picture number was defined as the eigenvalue of $\tilde{P}=-\partial \phi+\partial(i\kappa)$. For the physical states that are considered there, $P$ and $\tilde{P}$ agree since $\partial(i\kappa)$ has $0$ eigenvalue on these states.} In particular, we have
	\begin{equation} \label{eq:picture-vertex}
		\Phi=\phi \ e^{m\rho+in\sigma+ik_1 H^1+i k_2 H^2} D: \quad P(\Phi)=-m \ ,
	\end{equation}
	where $\phi$ is a field in the $\mathfrak{u}(1,1|2)_1$ free-field realisation, and $D$ is any combination of the derivatives of the hybrid fields. The physical states, in the topological description, are defined using a double cohomology expressed in eq.~(\ref{eq:physical}). There is a picture raising operator, $P_+$, that maps the physical states to the physical states and raises the picture number $P$. It is defined as \cite{Lust:1989tj}
	\begin{equation} \label{eq:picture-raising}
		P_+ = - G^+_0 (e^{-\rho-iH})_0 \ .
	\end{equation}
	Since different terms in $G^+$ have different eigenvalues under $P$, $P_+$ increases the eigenvalue of $P$ by either $1$ or $2$. However, we can define another picture number by
	\begin{equation}
		P^{\prime}=P+\frac{Y}{2} \ ,
	\end{equation}
	so that each term in $G^+$ has a fixed $P^{\prime}$ eigenvalue equal to $0$, and $P_+$ increases the $P^{\prime}$ eigenvalue by $1$. The statement that in the $n$-point functions we should have $\sum P_i=-2n$, is equivalent to $\sum P^{\prime}_i=-2n$, because we should have $\sum Y_i=0$, see \cite[Section~2]{Gaberdiel:2021njm}. Having this, when we are talking about the ``picture number'', we refer to the $P$ eigenvalue, or from eq.~(\ref{eq:picture-vertex}), just minus the coefficient of $\rho$ in the exponent.
	
	Finally, we discuss the picture numbers of the DDF operators and the physical states. In particular, we show that if $F$ is a DDF operator and $\Phi$ is a physical state, then in cohomology (i.e.\ up to BRST exact states) we have
	\begin{equation} \label{eq:picture-ddf-state}
		P_+(F_0 \Phi)(z)=F_0 (P_+ \Phi)(z)=(P_+ F)_0 \Phi(z) \ .
	\end{equation}
	Suppose we have one of the DDF operators $F$ (that in particular (anti-)commutes with $G^+_0$ and $\tilde{G}^+_0$) and a physical state $\Phi$ (which is annihilated by $G^+_0$ and $\tilde{G}^+_0$). Now we want to show that it does not matter in cohomology how one distributes the picture number between the DDF operator $F$ and the physical state $\Phi$. First consider $F_0 \Phi$. The state associated to it is $F_0 \Phi_0 \left|0\right>$ since $\Phi$ has world-sheet weight $0$ ($\left|0\right>$ is the vacuum on the world-sheet). We then have
	\begin{equation}
		P_+ (F_0 \Phi)(z) = V(G^+_0 \xi_0 F_0 \Phi_0 \left|0\right>,z) \ ,
	\end{equation}
	where $\xi=-e^{-\rho-iH}$, see eq.~(\ref{eq:picture-raising}). Now the state that corresponds to $F_0 (P_+ \Phi)(z)$ is
	\begin{equation}
		F_0 (P_+ \Phi)(z)=V(F_0 G^+_0 \xi_0 \Phi_0 \left|0\right>,z) \ .
	\end{equation}
	Since $F$ is a DDF operator, it (anti-)commutes with $G^+_0$ and $\tilde{G}^+_0$ possibly up to $Z$-terms which are zero on the physical states and can be ignored. Then we have
	\begin{equation}
		F_0 G^+_0 \xi_0 \Phi_0 \left|0\right> = \pm G^+_0 F_0 \xi_0 \Phi_0 \left|0\right> \ ,
	\end{equation}
	where $\pm$ stands for $F$ being bosonic ($+1$) or fermionic ($-1$). Now we use that for $\eta=\tilde{G}^+$ we have
	\begin{equation} \label{eq:xi-eta}
		\{\eta_n,\xi_m\}=\delta_{n+m,0} \ , 
	\end{equation}
	and insert $\{\eta_0,\xi_0\}=1$ right before $F_0$. This leads to
	\begin{equation}
		F_0 G^+_0 \xi_0 \Phi_0 \left|0\right> = \pm G^+_0 (\eta_0 \xi_0 + \xi_0 \eta_0 )F_0 \xi_0 \Phi_0 \left|0\right> \ .
	\end{equation}
	The first term is a BRST exact state because it has the form $G^+_0 \tilde{G}^+_0 \Phi^{\prime}$ and so can be ignored in cohomology. The second term gives
	\begin{equation}
		(\pm 1)^2 G^+_0 \xi_0 F_0 \eta_0 \xi_0 \Phi_0 \left|0\right> = G^+_0 \xi_0 F_0 \Phi_0 \left|0\right> \ ,
	\end{equation}
	where we have used that $\eta_0 \Phi=0$. This shows that in cohomology we have
	\begin{equation}
		P_+(F_0 \Phi)(z)=F_0 (P_+ \Phi)(z) \ . 
	\end{equation}
	Next consider the state that corresponds to $(P_+ F)_0 \Phi(z)$, which is
	\begin{equation}
		[G^+_0,(\xi F)_0] \Phi_0 \left|0 \right>= G^+_0 (\xi F)_0 \Phi_0 \left|0\right> \ ,
	\end{equation}
	where by $(\xi F)$ we mean the radial normal-ordering of $\xi$ with $F$. Let us insert $\{\eta_0,\xi_0\}=1$ before $(\xi F)_0$. We get
	\begin{equation}
		G^+_0 (\eta_0 \xi_0 + \xi_0 \eta_0 ) (\xi F)_0 \Phi_0 \left|0\right> \ ,
	\end{equation}
	where the first term is BRST exact and can be ignored in cohomology. Now consider the anti-commutator in eq.~(\ref{eq:xi-eta}). Since $\eta_0 \Phi=0$, only the zero mode of $\xi$ contributes because otherwise, $\eta_0$ anti-commutes with $\xi_n$ and $F_{-n}$ and kills $\Phi$. Indeed, all the modes of $F$ (anti-)commute with all the modes of $\eta=\tilde{G}^+$, or in other words, the integrand of the DDF operator $F$ has trivial OPE with $\tilde{G}^+$. So in the end, we get
	\begin{equation}
		G^+_0 \xi_0 \eta_0 \xi_0 F_0 \Phi_0 \left|0\right> = G^+_0 \xi_0 F_0 \Phi_0 \left|0\right> \ .
	\end{equation}
	This shows that eq.~(\ref{eq:picture-ddf-state}) holds in cohomology. It means that it does not matter how one divides the picture number between the DDF operator and the physical state.
	
	\section{BRST exactness of $\mathcal{L}_{-1} \label{app:exactness} \Omega^{w=1,P=-2}$}
	In this short appendix, we show that $\psi=\mathcal{L}_{-1} \Omega^{w=1,P=-2}$ is BRST exact up to $Z$-terms. In other words, as expected from the space-time theory, the translation of the vacuum is BRST exact. First notice that $\mathcal{L}_{-1}=J^+_0$, see eq.~(\ref{eq:j+-bosonised}), eq.~(\ref{eq:ddf-stress-tensor}) and \cite[eqs.~(A.2)]{Gaberdiel:2021njm}. We want to find $\phi$ such that $\psi=G^+_0 \phi$ and $\tilde{G}^+_0 \phi=0$, see eq.~(\ref{eq:n=2-g+}) and eq.~(\ref{eq:n=4-gtillde+}). This shows that $\psi=G^+_0 \tilde{G}^+_0 (\xi_0 \phi)$ where $\xi=-e^{-\rho-iH}$, and therefore $\psi$ is BRST exact (see eq.~(\ref{eq:xi-eta})). We first calculate
	\begin{equation}
		\psi=\mathcal{L}_{-1} \Omega^{w=1,P=-2}=J^+_0 \Omega^{w=1,P=-2}=e^{\phi_1+i\kappa_1+\phi_2} \partial(i\kappa_1) e^{2\rho+i\sigma+i H} \ .
	\end{equation}
	We make the following ansatz
	\begin{equation}
		\phi=\alpha \ e^{iq_2-iq_1} e^{2\phi_1+2i\kappa_1-i\kappa_2} e^{3\rho+i\sigma+i H} \partial(\rho+iH) + \beta \ e^{\phi_1+i\kappa_1+\phi_2} e^{2\rho+i H} \ .
	\end{equation}
	By a straightforward and direct computation, it can be seen that $\tilde{G}^+_0 \phi=0$. Moreover, it is also straightforward to calculate the action of $G^+_0$ on $\phi$. The result is
	\begin{equation}
		G^+_0 \phi = e^{\phi_1+i\kappa_1+\phi_2} e^{2\rho+i\sigma+iH} D^{\prime} \ ,
	\end{equation}
	where $D^{\prime}$ equals
	\begin{equation}
		D^{\prime} =\alpha[-\partial(2\rho+iH)-2\partial(\phi_1+i\kappa_1)+\partial(iq_1-iq_2)] + \beta [\partial(\phi_1+i\kappa_1+\phi_2) +\partial(2\rho+iH) ] \ .
	\end{equation}
	Setting $\beta=\alpha$ we get
	\begin{equation}
		D^{\prime\prime}=\alpha[-\partial(\phi_1+i\kappa_1)+\partial(iq_1-iq_2) +\partial \phi_2] \ .
	\end{equation}
	From eq.~(\ref{eq:z}), note that $Z$ equals
	\begin{equation}
		Z=U+V=\frac{1}{2}[-\partial \phi_1+\partial \phi_2] + \frac{1}{2}[\partial(iq_1)-\partial(iq_2)] \ .
	\end{equation}
	Now we set $\alpha=\beta=-1$, so we have
	\begin{equation}
		G^+_0 \phi= e^{\phi_1+i\kappa_1+\phi_2} \partial(i\kappa_1) e^{2\rho+i\sigma+iH} - 2 Z e^{\phi_1+i\kappa_1+\phi_2} e^{2\rho+i\sigma+iH} \ .
	\end{equation}
	Thus, $\mathcal{L}_{-1} \Omega^{w=1,P=-2}$ equals $G^+_0 \phi$ where $\tilde{G}^+_0 \phi=0$, up to a $Z$-term. This shows that it is BRST exact.

\end{document}